\documentclass[11pt,a4paper]{article}

\usepackage[T1]{fontenc}
\usepackage[utf8]{inputenc}

\usepackage{amsmath,amssymb,amsthm,mathtools}
\newtheorem{theorem}{Theorem}[section]
\newtheorem{proposition}[theorem]{Proposition}
\newtheorem{lemma}[theorem]{Lemma}
\newtheorem{corollary}[theorem]{Corollary}
\theoremstyle{definition}
\newtheorem{example}[theorem]{Example}
\newtheorem{assumption}[theorem]{Assumption}
\newtheorem{definition}[theorem]{Definition}
\theoremstyle{remark}
\newtheorem{remark}[theorem]{Remark}

\usepackage[margin=2cm]{geometry}
\usepackage{titlesec}

\usepackage{graphicx}
\usepackage{float}
\usepackage{caption}
\usepackage{booktabs}
\usepackage{tabularx}

\usepackage{algorithm}
\usepackage{algorithmic}

\usepackage[authordate, backend=biber, hypername=true]{biblatex-chicago}
\usepackage{xcolor}
\definecolor{db}{RGB}{0,0,139}
\usepackage[colorlinks=true,citecolor=db,linkcolor=db,urlcolor=blue,pdfstartview=FitH]{hyperref}

\usepackage{authblk}

\title{Data-Driven Dynamic Portfolio Selection with Vine Copula Simulation and Reinforcement Learning\thanks{This work is supported by the start-up fund of The Chinese University of Hong Kong, Shenzhen under Grant No. UDF004253. Most of this work was completed while G.M.J. Gelissen visited The Chinese University of Hong Kong in Shenzhen, China.}}

\author[1,2]{Gabriël M.J. Gelissen\thanks{%
Email: \href{mailto:g.gelissen@tum.de}{\texttt{g.gelissen@tum.de}}.}} %

\author[2]{Fengyi Yuan\thanks{Corresponding author. %
Email: \href{mailto:yuanfengyi@cuhk.edu.cn}{\texttt{yuanfengyi@cuhk.edu.cn}}.}} %

\affil[1]{Department of Mathematics, Technical University of Munich, Munich, Germany}

\affil[2]{School of Science and Engineering (SSE),
The Chinese University of Hong Kong, Shenzhen, China}

\date{}

\begin{document}

\hypersetup{
  pdftitle={Data-Driven Dynamic Portfolio Selection with Vine Copula Simulation and Reinforcement Learning},
  pdfauthor={Gabriel M.J. Gelissen and Fengyi Yuan},
  pdfsubject={Portfolio Optimisation, Vine Copulas, Deep Reinforcement Learning, Dynamic Dependence, Conditional Value-at-Risk (CVaR), Synthetic Data Generation}
}

\maketitle
\vspace{-3em}

\begin{abstract}
\noindent We develop a data-driven approach to dynamic portfolio selection that combines vine-copula simulation with reinforcement learning. The investor has constant relative risk aversion and maximises expected terminal utility with a running penalty on the conditional value-at-risk (CVaR) of one-period losses. We estimate a conditional vine model from historical returns and use it to generate joint scenarios for CVaR evaluation and conditional paths for training a recurrent portfolio policy. Under uniform local derivative bounds, we propagate edgewise pair-copula parameter errors through the conditional transforms to obtain power-law stability bounds in Hellinger distance and total variation. In a finite-horizon investment model, we combine these estimates with a marginal and structural approximation budget to bound policy values, optimal values and the true-market loss of policies selected under a fitted model, with separate simulation and numerical optimisation terms. On a seven-asset panel comprising US and Chinese equities and gold, the selected ensemble of ten trained policies achieves a compound annual growth rate of 34.13\% and a Sharpe ratio of 2.31 over 22 complete monthly holding periods, exceeding the corresponding values of six classical and vine-based benchmarks. Combining simulated and historical training data reduces monthly target-to-target turnover by 41.5\% relative to training on historical data alone, while maintaining comparable growth.
\end{abstract}


\section{Introduction}
\label{sec:introduction}

Portfolio choice is central to modern finance theory. Foundational studies include Markowitz's static portfolio selection \autocite{markowitzPortfolioSelection1952b} and Merton's continuous-time portfolio problem \autocite{mertonOptimumConsumptionPortfolio1971}. In the dynamic portfolio choice problem, an investor must allocate wealth across assets while responding to new information and accounting for the effects of current positions on future wealth and risk exposure. Implementing a dynamic strategy from historical data requires estimating the return distribution and learning how allocations should respond to changing market and portfolio states. With several risky assets, the dependence structure determines how individual risks combine and how diversification performs when losses occur jointly. A data-driven approach to risk-adjusted portfolio selection must therefore connect conditional dependence modelling with the investor's sequential decisions.

In this paper, we develop a data-driven approach to dynamic portfolio selection that combines vine-copula simulation with reinforcement learning. The investor has constant relative risk aversion (CRRA) and maximises expected terminal utility with a running penalty on the conditional value-at-risk (CVaR) of one-period portfolio losses. Vine copulas provide a convenient tool for modelling dynamic multivariate return distributions: conditional marginal models describe individual assets, while pair copulas with history-dependent parameters describe their evolving dependence and permit recursive conditional sampling. Their pair-copula factorisation also makes quantitative error propagation tractable, allowing us to trace local dependence-parameter errors through the conditional transforms to the joint return law and, subsequently, to investment values and policy performance. We estimate conditional marginal distributions and a vine copula from historical returns. The fitted model supplies joint scenarios for evaluating the CVaR of candidate allocations and conditional return paths for learning a recurrent portfolio policy. Reinforcement learning uses the resulting experience to approximate investment decisions under portfolio constraints and trading costs.

Conditional simulation plays a central role in the approach. Historical observations provide a finite collection of market paths, while policy learning requires evaluating allocations across evolving combinations of returns, wealth and positions. Updating the fitted conditional model along simulated paths supplies additional training experience under the estimated return dynamics. This construction connects the estimated joint return law to both current risk evaluation and future investment value. We quantify how model and numerical approximation errors affect policy performance, and evaluate the method on financial market data.

\subsection{Related work}

The use of estimated return models connects our approach to the literature on model misspecification in portfolio optimisation. \textcite{demiguelOptimalNaiveDiversification2009} highlight the economic importance of estimation error in comparisons of optimised portfolios with equal weighting. Distributionally robust formulations account for uncertainty about the return law by considering adverse distributions near a reference model. In a one-period market, \textcite{oblojDistributionallyRobustPortfolio2021} derive first-order sensitivities of investment values, optimal portfolios and marginal utility prices under Wasserstein model uncertainty. These results motivate assessing a fitted return model through its implications for investment values and decisions.

For dynamic investment, the conditional evolution of returns is also relevant to the quality of the fitted model. \textcite{backhoff-veraguasAdaptedWassersteinDistances2020} establish stability results for hedging and utility maximisation using adapted Wasserstein distances, which account for the temporal information structure. \textcite{bartlSensitivityMultiperiod2023} obtain first-order approximations for multiperiod optimisation values under perturbations measured by this distance. In our framework, the estimated conditional law determines scenario-based risk measures and simulated investment paths. We use the vine hierarchy to quantify approximation and fitting errors in this conditional law and their effects on the performance of portfolio policies.

Vine copulas provide a flexible representation of the conditional joint return distribution. Copulas separate marginal distributions from dependence, while pair-copula constructions express a multivariate density through bivariate copulas and recursively computed conditional distributions \autocite{joeMultivariateModelsDependence1997,bedfordProbabilityDensityDecomposition2001a,bedfordVinesNewGraphical2002a,aasPaircopulaConstructionsMultiple2009}. The same decomposition supports likelihood evaluation and sampling. Once the marginal dynamics and dependence parameters have been estimated, conditional sampling generates joint return scenarios, and updating these models sequentially produces return paths. These two operations supply the risk evaluations and state transitions used by the portfolio learning algorithm.

The financial use of conditional copulas is well established. \textcite{bernardiPortfolioOptimisationFlexible2018} relate dynamic dependence and higher conditional moments to utility-based portfolio allocation. \textcite{sahamkhadamPortfolioOptimizationBased2023a} use vine-based return forecasts to compare allocation criteria including minimum CVaR. These studies connect flexible joint forecasts with portfolio risk measurement and optimisation. We use a conditional vine model as the simulation environment for a finite-horizon investment problem. The learned policy responds to market and portfolio histories, while the simulated returns determine how its current allocations affect wealth, positions and subsequent decisions.

Reinforcement learning provides a numerical approach to sequential portfolio choice. \textcite{moodyLearningTradeDirect2001} develop direct reinforcement learning for trading with transaction costs, and \textcite{almahdiAdaptivePortfolioTrading2017} study recurrent portfolio policies under a drawdown-based risk criterion. \textcite{coacheConditionallyElicitable2023} construct actor--critic methods for time-consistent dynamic spectral risk measures and apply them to portfolio allocation. In our formulation, conditional CVaR enters as a one-period risk charge, and continuation values use ordinary conditional expectation. The representation of \textcite{rockafellarOptimizationConditionalValueatRisk2000,rockafellarConditionalValueatriskGeneral2002} permits scenario evaluation of this risk charge.

Conditional generative models offer a way to supply the investment paths required for learning. \textcite{aghapourDynamicDataGeneration2026} develop conditional score-based diffusion models for dynamic data generation, establish stability results for dynamic mean--variance portfolio selection and train portfolio policies in the generated environment. Their work links the accuracy of conditional path generation to the quality of dynamic investment decisions. Our approach develops this connection for vine-copula simulation and a CRRA--CVaR investment objective. The pair-copula representation supports both simulation of the learning environment and an explicit analysis of the resulting investment values and policy performance.

\subsection{Results and contributions}

We provide a theoretical analysis in a finite-horizon investment model using an auxiliary law with the fitted conditional margins, selected simplified-vine structure and fixed pair-copula family on each edge. Under uniform local derivative bounds, Theorem~\ref{thm:vine_density_stability} propagates edgewise parameter errors along the conditional-input recursion and compares the auxiliary and fitted laws in Hellinger distance and total variation. Balancing the local propagation bound with the probability of boundary regions yields a stability bound of order $\varepsilon^{1/(2d-1)}$ for parameter tolerance $\varepsilon$ in dimension $d$. The result covers Student-$t$, Gaussian and Clayton pair copulas on the specified compact parameter domains. Lemma~\ref{lem:vine_model_error} combines these estimates with a uniform relative-entropy budget between the true and auxiliary laws to account for marginal and structural misspecification.

Theorem~\ref{thm:vine_policy_performance} converts these joint-law bounds into uniform policy-value and optimal-value bounds for the CRRA--CVaR investment problem on a common policy class with a uniform second-moment bound on terminal utility. The terminal-utility moment budget determines the horizon dependence of the performance bound, with an $O(T^{3/2})$ coefficient for logarithmic utility when the model constants are fixed. Theorem~\ref{thm:vine_policy_transfer} then controls the true-market loss of a policy selected under the fitted model. Corollary~\ref{cor:vine_learning} separates marginal and structural approximation, pair-copula parameter error, simulation error and numerical optimisation under explicit uniform-error and optimisation-gap conditions. This analysis connects approximation of the conditional simulation environment to the performance of the resulting dynamic investment decisions.

We implement the approach with a neural Student-$t$ D-vine and a recurrent twin-delayed deep deterministic policy-gradient (TD3) algorithm. A constrained action map enforces the portfolio mandate, and the reward incorporates utility increments, trading costs and a scenario-CVaR charge. Training combines simulated and historical episodes before historical fine-tuning. Averaging the target weights of independently trained policies gives the ensemble strategy used in the empirical comparison.

The empirical study evaluates our framework on a seven-asset panel comprising US and Chinese equities and gold. Over 22 complete monthly holding periods, the selected ensemble of ten trained policies records a compound annual growth rate of 34.13\% and a Sharpe ratio of 2.31, exceeding the corresponding values of six classical and vine-based benchmarks. Its monthly target-to-target turnover is approximately one third of that of the static and dynamic neural-vine optimisers. Relative to training on historical data alone, mixed training maintains comparable ensemble growth while reducing turnover by 41.5\%. These comparisons describe the observed performance of the selected ensemble on this evaluation sample.

Our main theoretical contribution is a quantitative analysis linking local pair-copula parameter errors to dynamic investment values and policy performance. We summarise our contributions as follows.
\begin{itemize}
    \item We combine a dynamic vine-copula model with reinforcement learning to construct portfolio strategies under a finite-horizon CRRA--CVaR objective. The same fitted conditional return law supplies scenarios for conditional CVaR evaluation and training paths for a recurrent policy under portfolio constraints and transaction costs.
    \item To the best of our knowledge, we provide the first quantitative analysis that propagates pair-copula parameter errors through the recursive conditional inputs of a vine to control dynamic CRRA--CVaR investment values and policy-transfer losses. Under our local derivative assumptions, boundary localisation yields explicit Hellinger and total-variation bounds for the joint return law while allowing pair-copula densities with boundary singularities. Under our moment and approximation conditions, these estimates combine with the fitted-model optimisation gap to control the true-market performance loss of a selected policy.
    \item We provide empirical evidence of the benefits of vine-based policy learning on a seven-asset panel. Over the 22-period evaluation sample, the selected ensemble achieves higher observed growth and a higher Sharpe ratio than six classical and vine-based benchmarks. Mixing simulated and historical training paths reduces monthly target-to-target turnover by 41.5\% relative to historical-only training while maintaining comparable growth.
\end{itemize}

\paragraph{Organisation of the paper.}
The remainder of the paper is organised as follows. Section~\ref{sec:problem_formulation} formulates the market, conditional vine model and investment objective. Section~\ref{sec:dependence_robustness} develops the pair-copula parameter, dynamic-value and policy-performance bounds. Section~\ref{sec:setup_implementation} describes the numerical implementation and training procedure. Section~\ref{sec:empirical_study} presents the data and empirical results. Section~\ref{sec:conclusion} concludes. The \hyperref[sec:appendix]{Appendix} contains implementation details, supplementary experiments and complete proofs.

\section{Problem formulation}
\label{sec:problem_formulation}

We formulate a finite-horizon investment problem in which the conditional joint return law determines both a portfolio's current tail-risk charge and its future investment opportunities. We compare the true market with a fitted conditional vine model, keeping the investor's information, portfolio constraints and physical wealth dynamics fixed. We separate marginal and structural approximation error from pair-copula parameter errors within the selected vine, and trace their effects through the joint return law to investment values and the performance of fitted-model policies.

\subsection{Market dynamics and admissible strategies}
\label{subsec:market_dynamics}

We consider a discrete-time financial market over a finite horizon $T$, with a cash account and $d$ risky assets. Trading takes place at dates $t=0,\ldots,T-1$. We work on a filtered probability space $(\Omega,\mathcal F,\mathbb F,\mathbb P)$, where $\mathbb F=(\mathcal F_t)_{t=0}^T$ and $\mathbb P$ is the true market law. All prices and wealth are expressed in units of the cash account. Thus its discounted price is $S_t^0=1$, and $S_t^i>0$ denotes the discounted price of risky asset $i$. The vector of one-period discounted gross returns is $G_{t+1}=(G_{1,t+1},\ldots,G_{d,t+1})$, where
\begin{equation*}
 G_{i,t+1}=\frac{S_{t+1}^i}{S_t^i},\qquad i=1,\ldots,d,
 \qquad G_{t+1}\in\mathcal G\subset(0,\infty)^d.
\end{equation*}
The investor observes the market history $H_t=(H_0,G_1,\ldots,G_t)$, and $\mathcal F_t=\sigma(H_t)$. Here $H_0$ records the initial asset prices and any presample observations, so the price processes are $\mathbb F$-adapted. Under $\mathbb P$, the conditional distribution of $G_{t+1}$ given $\mathcal F_t$ is $q_t(dg\mid H_t)$. We write $\mathbb E$ for expectation under $\mathbb P$.

Let $W_t$ be the investor's discounted wealth immediately before trading at date $t$, with initial wealth $W_0>0$, and set $X_t=W_t/W_0$. We denote the risky-asset proportions before rebalancing by $b_t\in\mathbb R^d$. At date $t$, the investor chooses an $\mathcal F_t$-measurable allocation $a_t\in\mathbb R^d$, investing $W_ta_{i,t}$ in asset $i$. Both $b_t$ and $a_t$ are measured relative to the same pretrade wealth $W_t$. In particular, the number of shares held over $(t,t+1]$ is $\varphi_t^i=W_ta_{i,t}/S_t^i$.

Rebalancing incurs a transaction cost $W_tc_t(b_t,a_t)$, paid from the cash account, where $c_t\geq0$ is the cost per unit of pretrade wealth. Proportional transaction costs with rate $\kappa_t\geq0$ correspond to $c_t(b,a)=\kappa_t\|a-b\|_1$. More general continuous convex cost functions are also allowed. After trading and paying costs, the cash holding is $W_t[1-\mathbf1^\top a_t-c_t(b_t,a_t)]$. The self-financing condition therefore gives
\vspace{-10pt}
\begin{align*}
 W_{t+1}
 &=W_t\bigl[1-\mathbf1^\top a_t-c_t(b_t,a_t)\bigr]
   +\sum_{i=1}^d\varphi_t^iS_{t+1}^i\\
 &=W_t\bigl[1+a_t^\top(G_{t+1}-\mathbf1)-c_t(b_t,a_t)\bigr].
\end{align*}
Here $\mathbf1$ is the vector of ones.

For the purpose of dynamic programming and reinforcement learning, we also model the portfolio selection through the lens of {\it Markov Decision Processes}. More specifically, we use the state $s_t=(H_t,X_t,b_t)$. Given a state $s=(H,X,b)$, allocation $a$ and return $g$, define the net wealth growth factor $K_t$ and the state transition $\Phi_t$ by
\begin{align}
 K_t(s,a,g)&=1+a^\top(g-\mathbf1)-c_t(b,a), \notag\\
 \Phi_t(s,a,g)&=\left((H,g),\;XK_t(s,a,g),\;
              \frac{a\odot g}{K_t(s,a,g)}\right).
 \label{eq:ideal_market_dynamics}
\end{align}
Here $\odot$ denotes componentwise multiplication. Consequently $s_{t+1}=\Phi_t(s_t,a_t,G_{t+1})$. The last component of $\Phi_t$ gives the risky-asset proportions at the next trading date, after the intervening price changes. The corresponding cash proportion is $[1-\mathbf1^\top a_t-c_t(b_t,a_t)]/K_t(s_t,a_t,G_{t+1})$, so the risky and cash proportions sum to one. We fix the initial state $s_0=(H_0,1,b_0)$ throughout.

Write $A_t(s)\subset\mathbb R^d$ for the set of feasible portfolio allocations at state $s$. Portfolio constraints may include individual position limits, a gross-exposure bound $\|a\|_1\leq\Lambda$, and bounds on net risky exposure and cash holdings after costs. We use the states reachable from $s_0$ as our state spaces, defined recursively by
\begin{equation}
 \mathcal S_0=\{s_0\},\qquad
 \mathcal S_{t+1}
 =\{\Phi_t(s,a,g):s\in\mathcal S_t,\ a\in A_t(s),\ g\in\mathcal G\}.
 \label{eq:reachable_state_spaces}
\end{equation}
Equip $\mathcal S_t\subset(\{H_0\}\times\mathcal G^t)\times(0,\infty)\times\mathbb R^d$ with the product metric. The solvency condition below makes this recursion well defined. We allow every return in $\mathcal G$ in this construction, so the state spaces do not depend on the choice of conditional return kernels.

\begin{assumption}
\label{ass:market}
The horizon $T$ is finite, the initial state $s_0=(H_0,1,b_0)$ is fixed, and the common return domain $\mathcal G\subset(0,\infty)^d$ is a nonempty Polish subspace. At every date $t<T$, the following conditions hold.
\begin{enumerate}
 \item The cost function $c_t$ is nonnegative and continuous. On $\mathcal S_t$, the feasible correspondence $A_t$ is nonempty, continuous and compact-valued. There is a constant $\Lambda<\infty$, independent of $t$ and $s$, such that $\|a\|_1\leq\Lambda$ for every $a\in A_t(s)$.
 \item There is a constant $\underline k>0$, independent of $t$, $s$ and $a$, such that every $s\in\mathcal S_t$ and $a\in A_t(s)$ satisfy
 \begin{equation}
  \inf_{g\in\mathcal G}K_t(s,a,g)\geq\underline k.
  \label{eq:ideal_solvency}
 \end{equation}
 \item The exogenous return kernel $q_t(dg\mid H)$ is a probability kernel on $\mathcal G$ and is weakly continuous in $H\in\{H_0\}\times\mathcal G^t$.
 \item There is a constant $B<\infty$, independent of $t$, $s$ and $a$, such that every $s=(H,X,b)\in\mathcal S_t$ and $a\in A_t(s)$ satisfy
 \begin{equation*}
  \int_{\mathcal G}K_t(s,a,g)q_t(dg\mid H)\leq B.
 \end{equation*}
\end{enumerate}
\end{assumption}

\begin{definition}
\label{def:admissible_policy}
An admissible portfolio policy $\pi=(\pi_t)_{t=0}^{T-1}$ consists of Borel measurable maps $\pi_t:\mathcal S_t\to\mathbb R^d$ satisfying $\pi_t(s)\in A_t(s)$ for every $s\in\mathcal S_t$. Its implemented allocation is $a_t=\pi_t(s_t)$, with $s_{t+1}=\Phi_t(s_t,a_t,G_{t+1})$ and initial state $s_0$.
\end{definition}

Since the state contains the full observed history, Definition~\ref{def:admissible_policy} includes adapted deterministic investment rules with history dependence. The resulting wealth remains strictly positive by Equation~\eqref{eq:ideal_solvency}. We next establish regularity of the reachable state spaces from the primitive conditions.

\begin{lemma}
\label{lem:reachable_regularity}
Under Assumption~\ref{ass:market}, each $\mathcal S_t$ is nonempty and relatively closed in $(\{H_0\}\times\mathcal G^t)\times(0,\infty)\times\mathbb R^d$, and hence is a Polish space. The transition $\Phi_t$ is continuous on the feasible state--action--return domain. Every reachable state $s=(H,X,b)\in\mathcal S_t$ satisfies
\begin{equation}
 X\geq\underline k^{\,t}>0.
 \label{eq:market_state_bounds}
\end{equation}
\end{lemma}

\begin{proof}
See \hyperref[app:reachable_regularity_proof]{Appendix~\ref*{app:problem_formulation_proofs}}.
\end{proof}

\subsection{Conditional vine models and dependence misspecification}
\label{subsec:conditional_vine}

Fix a trading date $t$ and an observed market history $H$. Throughout the construction below, all distributions are conditional on $H$, and we write $G=(G_1,\ldots,G_d)$ for $G_{t+1}$, suppressing the date and history from the notation. Suppose that $G$ has joint density $q$, with marginal distribution functions $F_i$ and densities $f_i$. The return density therefore factors as
\begin{equation}
 q(g)=c\bigl(F_1(g_1),\ldots,F_d(g_d)\bigr)
       \prod_{i=1}^d f_i(g_i).
 \label{eq:conditional_market_law}
\end{equation}
A regular vine further decomposes $c$ into bivariate copula densities arranged in $d-1$ tree levels. Let $\mathcal E_k$ be the edge set at level $k$. Each edge $e=(i_e,j_e;D_e)$ specifies two asset indices $i_e,j_e$ and a set $D_e$ indexing the other contemporaneous returns on which we condition. At $u_i=F_i(g_i)$, the decomposition takes the form
\begin{equation}
 c(u)
 =\prod_{k=1}^{d-1}\prod_{e\in\mathcal E_k}
 c_e\bigl(u_{i_e\mid D_e},u_{j_e\mid D_e}\mid g_{D_e}\bigr).
 \label{eq:conditional_vine_factorisation}
\end{equation}
Here $c_e(\cdot,\cdot\mid g_{D_e})$ is the copula density of $(G_{i_e},G_{j_e})$ conditional on $G_{D_e}=g_{D_e}$. Its two arguments are the conditional probability integral transforms
\begin{equation*}
 u_{i\mid D}=F_{i\mid D}(g_i\mid g_D),
 \qquad u_{i\mid\varnothing}=u_i,
\end{equation*}
where $F_{i\mid D}$ is the conditional distribution function of $G_i$ given $G_D$. Thus each factor describes the dependence remaining between two returns after conditioning on the returns indexed by $D_e$. All these distributions also remain conditional on the fixed market history $H$.

The two products in \eqref{eq:conditional_vine_factorisation} show how these pairs form the \emph{vine tree hierarchy}. At level $k$, there are $d-k$ factors, each with a conditioning set of size $k-1$. The first level therefore contains $d-1$ pair copulas conditional only on $H$. The second contains $d-2$ pair copulas, each additionally conditional on one contemporaneous return. The final level contains a single pair copula conditional on the other $d-2$ returns. Multiplying over edges and then over levels recovers the full joint copula density \autocite{bedfordVinesNewGraphical2002a,aasPaircopulaConstructionsMultiple2009}.

The trees determine which conditional pairs appear in this product. Formally, a regular vine consists of trees $T_k=(N_k,\mathcal E_k)$, where $N_1=\{1,\ldots,d\}$ and $N_{k+1}=\mathcal E_k$. An edge at one level becomes a node at the next. The proximity condition requires two nodes joined in $T_{k+1}$ to represent edges of $T_k$ that share a node. This condition ensures that the conditional arguments required at each level can be constructed from the preceding levels. For any joint law with a density, a full regular vine gives a factorisation of the form \eqref{eq:conditional_vine_factorisation}, up to null sets.

For the fitted model, we use simplified vines, in which each pair-copula density is independent of $g_{D_e}$ at fixed copula arguments $(v,w)$ \autocite{haffSimplifiedPairCopula2010}. Under this assumption we write the factor as $c_e(v,w)$, while its conditional marginal arguments still depend on $g_{D_e}$. The true market law may have different margins and need not admit a simplified representation on the selected vine. Section~\ref{sec:dependence_robustness} includes these discrepancies through an auxiliary model and an abstract approximation error.

\begin{remark}
We take a three-asset D-vine as an example. It can be decomposed into first-tree edges $(1,2)$, $(2,3)$ and the second-tree edge $(1,3;\{2\})$. Under the simplifying assumption, its copula density is
\begin{equation*}
 c(u_1,u_2,u_3)
 =c_{12}(u_1,u_2)c_{23}(u_2,u_3)
  c_{13\mid2}(u_{1\mid2},u_{3\mid2}).
\end{equation*}
The first two factors specify the joint laws of $(G_1,G_2)$ and $(G_2,G_3)$ once the margins are given. They also determine $u_{1\mid2}$ and $u_{3\mid2}$, which enter the last factor to specify the conditional dependence between $G_1$ and $G_3$ given $G_2$. Changing this last factor preserves those two first-tree joint laws, while the joint law of $(G_1,G_3)$ may change. Section~\ref{subsec:conditional_law_estimation} uses the corresponding D-vine construction for $d$ assets.
\end{remark}

\begin{remark}
The vine decomposition provides a recursive method for simulating joint return scenarios. Consider a simplified-vine law with marginal distribution functions $F_i$ and pair-copula densities $c_e$. For an edge $e=(i,j;D)$, the $h$-function computes a conditional distribution from its pair-copula density:
\begin{align*}
 h_e(v\mid w)&=\int_0^v c_e(z,w)\,dz,\\
 u_{i\mid D\cup\{j\}}
 &=h_e(u_{i\mid D}\mid u_{j\mid D}).
\end{align*}
The reverse orientation gives $u_{j\mid D\cup\{i\}}$. Applying these functions successively constructs the conditional distributions needed for sampling. After relabelling the assets in an order compatible with the vine, draw independent $Z_1,\ldots,Z_d\sim\mathcal U(0,1)$ and set
\begin{align*}
 \widetilde G_1&=F_1^{-1}(Z_1),\\
 \widetilde G_i&=F_{i\mid1:i-1}^{-1}
 \bigl(Z_i\mid\widetilde G_{1:i-1}\bigr),\qquad i=2,\ldots,d.
\end{align*}
Here $1:i-1=\{1,\ldots,i-1\}$, and the generalised inverse is taken in the distribution-function argument. This inverse Rosenblatt construction samples the joint law specified by these margins and pair copulas \autocite{aasPaircopulaConstructionsMultiple2009}. Independent repetitions at a fixed history give joint scenarios for evaluating portfolio losses and conditional CVaR. Updating the history after each simulated return and using the next conditional law instead produces paths for dynamic policy training.
\end{remark}

We study \emph{dependence misspecification} by comparing the true conditional return law $q_t(\cdot\mid H)$ with its fitted approximation $\widehat q_t(\cdot\mid H)$. The fitted kernels define a market law $\widehat{\mathbb P}$ on the same filtered measurable space, with expectation denoted by $\widehat{\mathbb E}$. Throughout, a hat distinguishes a fitted-model quantity from its true-market counterpart. The fitted model uses the return domain, initial state, state spaces, portfolio constraints, cost functions and state transition specified in Section~\ref{subsec:market_dynamics}. We impose Assumption~\ref{ass:market} on the fitted model with $\widehat q_t$ in place of $q_t$, using the same constants $\underline k$, $B$ and feasible action sets. In particular, $\widehat q_t$ is weakly continuous in $H$ and satisfies the same conditional first-moment bound on $K_t$.

To isolate pair-copula parameter errors, Section~\ref{subsec:vine_approximation} introduces an auxiliary conditional law $q_t^\circ(\cdot\mid H)$ with the fitted margins and selected simplified-vine structure. On each edge, the auxiliary and fitted models use the same fixed parametric copula family, while different edges may use different families. Their parameters may vary with the date and observed market history within fixed parameter domains. The relative-entropy budget $\zeta$ between $q_t$ and $q_t^\circ$ accounts for the fitted margins and the expressive restrictions of the selected families, vine structure and simplifying assumption. The comparison between $q_t^\circ$ and $\widehat q_t$ then propagates their edgewise parameter differences through each model's own conditional transforms. Section~\ref{sec:dependence_robustness} combines these two errors to bound dynamic optimal values and the true-market performance of fitted-model policies.

\subsection{CVaR-adjusted rewards and the investment objective}
\label{subsec:cvar_objective}

The one-period loss per unit of pretrade wealth is $\ell_t(s,a,g)=1-K_t(s,a,g)$. Fix a confidence level $\alpha\in(0,1)$ and write $p=1-\alpha$. We use the standard upper-tail conditional value-at-risk of this signed loss. For each state and action, its threshold representation is
\begin{equation}
 \mathcal C_t(s,a)
 =\min_{v\in\mathbb R}\left\{v+
 \frac{1}{p}\int_{\mathcal G}(\ell_t(s,a,g)-v)_+q_t(dg\mid H)\right\}.
 \label{eq:conditional_cvar}
\end{equation}
This representation follows \textcite{rockafellarOptimizationConditionalValueatRisk2000,rockafellarConditionalValueatriskGeneral2002}. Replacing $q_t$ with $\widehat q_t$ defines the fitted-model charge $\widehat{\mathcal C}_t(s,a)$. For signed losses, the charge can be negative when the upper loss tail corresponds to gains.
Assumption~\ref{ass:market} gives the uniform bounds
\begin{equation*}
 1-B\leq\mathcal C_t(s,a),\widehat{\mathcal C}_t(s,a)
 \leq1-\underline k.
\end{equation*}

For risk-aversion parameter $\gamma_U>0$, set
\begin{equation}
 U_{\gamma_U}(X)=
 \begin{cases}
 X^{1-\gamma_U}/(1-\gamma_U),&\gamma_U\ne1,\\
 \log X,&\gamma_U=1,
 \end{cases}
 \qquad
 u_t(s,a,g)=U_{\gamma_U}(XK_t(s,a,g))-U_{\gamma_U}(X).
\end{equation}
For $\lambda>0$, the realised reward and its conditional mean are
\begin{align}
 r_t(s,a,g)&=u_t(s,a,g)-\lambda\mathcal C_t(s,a),\notag\\
 \bar r_t(s,a)&=\int u_t(s,a,g)q_t(dg\mid H)
                  -\lambda\mathcal C_t(s,a).
 \label{eq:theoretical_reward}
\end{align}
The investor optimises over the admissible policies in Definition~\ref{def:admissible_policy}:
\begin{equation}
 J(\pi;s_0)=\mathbb E^\pi\!\left[
 \sum_{t=0}^{T-1}r_t(s_t,\pi_t(s_t),G_{t+1})\,\middle|\,s_0\right],
 \qquad V_0(s_0)=\sup_\pi J(\pi;s_0).
 \label{eq:investment_objective}
\end{equation}
The fitted-model rewards $\widehat r_t$ and $\widehat{\bar r}_t$, objective $\widehat J$, and optimal value $\widehat V_0$ are defined by the same formulas using $\widehat q_t$, $\widehat{\mathcal C}_t$, and $\widehat{\mathbb E}^{\pi}$. The utility increments telescope to terminal utility minus initial utility. The remaining term is the expected sum of conditional risk charges along the investment path. Subsequent values are evaluated by ordinary conditional expectation. Consequently, Equation~\eqref{eq:investment_objective} specifies an additive investment criterion with local CVaR charges, distinct from a nested dynamic risk measure or the CVaR of total terminal loss. For dynamic risk criteria built from conditional risk measures, see \textcite{coacheConditionallyElicitable2023}. 

\subsection{Dynamic programming}
\label{subsec:dynamic_programming}

For a fixed policy, let $V_t^{\pi}(s)$ be its remaining expected reward from $s$ at date $t$. The terminal values are zero, and backward induction gives
\begin{align}
 Q_t^{\pi}(s,a)
 &=\int\bigl[u_t(s,a,g)+V_{t+1}^{\pi}(\Phi_t(s,a,g))\bigr]
           q_t(dg\mid H)-\lambda\mathcal C_t(s,a),\notag\\
 V_t^{\pi}(s)&=Q_t^{\pi}(s,\pi_t(s)).
 \label{eq:fixed_policy_bellman}
\end{align}
For optimal control, define
\begin{align}
 Q_t(s,a)
 &=\int\bigl[u_t(s,a,g)+V_{t+1}(\Phi_t(s,a,g))\bigr]
           q_t(dg\mid H)-\lambda\mathcal C_t(s,a),\notag\\
 V_t(s)&=\max_{a\in A_t(s)}Q_t(s,a),
 \qquad V_T=0.
 \label{eq:optimal_bellman}
\end{align}
The fitted-model functions $\widehat Q_t^{\pi}$, $\widehat V_t^{\pi}$, $\widehat Q_t$, and $\widehat V_t$ satisfy the corresponding recursions with $\widehat q_t$ and $\widehat{\mathcal C}_t$.

\begin{proposition}
\label{prop:dynamic_programming}
Under Assumption~\ref{ass:market}, the value functions in Equation~\eqref{eq:optimal_bellman} are finite and continuous in $s$. There are Borel measurable maximisers $\pi_t^*(s)$, and the resulting policy attains Equation~\eqref{eq:investment_objective}. Every admissible policy has finite Borel measurable value functions satisfying Equation~\eqref{eq:fixed_policy_bellman}. These value functions are continuous when the policy is continuous. The same conclusions hold for the fitted model when it satisfies Assumption~\ref{ass:market}, with Borel measurable maximisers $\widehat\pi_t^*(s)$.
\end{proposition}

\begin{proof}
See \hyperref[app:dp_proof]{Appendix~\ref*{app:problem_formulation_proofs}}.
\end{proof}

The same conditional law enters both terms of Equation~\eqref{eq:optimal_bellman}: it evaluates the current CVaR charge and propagates the future state. A vine simulator therefore has two different purposes in our approach. Joint return scenarios evaluate candidate allocations under the risk charge, while coherent conditional paths supply transitions for continuation-value learning. Section~\ref{sec:dependence_robustness} quantifies how errors in that shared dependence law affect investment values and policy performance.

\section{Portfolio performance under dependence misspecification}
\label{sec:dependence_robustness}

We study how parameter errors in the pair copulas of a selected vine affect dynamic portfolio selection, allowing the marginal specification and the vine model class to be misspecified. An auxiliary model with the fitted margins and selected structure separates these sources of error. We impose a uniform tolerance $\varepsilon$ on the parameters of each pair copula and an approximation budget $\zeta$ between the true and auxiliary laws. Under a uniform local derivative bound, we propagate the parameter errors through the conditional transforms and obtain Hellinger and total-variation bounds for the joint return laws. These bounds control investment values and the performance of a policy selected under the fitted model. The true and fitted markets share the transition rule in Equation~\eqref{eq:ideal_market_dynamics}, the feasible policy class, and the form of the objective in Equation~\eqref{eq:investment_objective}. Appendix~\ref{app:dependence_results_proofs} provides the proofs.

\subsection{From pair-copula approximation to the joint return law}
\label{subsec:vine_approximation}

Fix a fitted model $\widehat q_t(\cdot\mid H)$ with conditional margins $\widehat F_{i,t}(\cdot\mid H)$ and a selected simplified regular vine $\widehat{\mathcal V}$ on $d\ge2$ assets. The structure is fixed across dates and histories, with edge set $\mathcal E=\bigcup_{k=1}^{d-1}\mathcal E_k$. The fitted margins have densities supported on common intervals $I_i\subset(0,\infty)$, with $\mathcal G=\prod_{i=1}^d I_i$. Each density is positive and continuous on the interior of its interval. We introduce an auxiliary conditional probability kernel $q_t^\circ(\cdot\mid H)$, measurable in $H$, with these same margins and structure. Its joint copula density $c_t^\circ$ and the fitted density $\widehat c_t$ give, for Lebesgue-almost every $g\in\mathcal G$,
\begin{align}
 q_t^\circ(g\mid H)
 &=c_t^\circ\bigl(\widehat F_{1,t}(g_1\mid H),\ldots,
                  \widehat F_{d,t}(g_d\mid H)\mid H\bigr)
   \prod_{i=1}^d\widehat f_{i,t}(g_i\mid H),\notag\\
 \widehat q_t(g\mid H)
 &=\widehat c_t\bigl(\widehat F_{1,t}(g_1\mid H),\ldots,
                  \widehat F_{d,t}(g_d\mid H)\mid H\bigr)
   \prod_{i=1}^d\widehat f_{i,t}(g_i\mid H).
 \label{eq:vine_reference_model}
\end{align}
We use the same symbol for a conditional kernel and its density, with $q_t^\circ(dg\mid H)=q_t^\circ(g\mid H)\,dg$ and likewise for the fitted law. The approximation budget below accounts for differences between the true kernel $q_t$ and this selected model class.

For each edge $e$, fix a parameter family $c_e(u,v;\theta_e)$ with $\theta_e\in\Theta_e\subset\mathbb R^{p_e}$ and write
\begin{equation*}
 c^\circ_{e,t}(u,v\mid H)=c_e(u,v;\theta^\circ_{e,t}(H)),
 \qquad
 \widehat c_{e,t}(u,v\mid H)=c_e(u,v;\widehat\theta_{e,t}(H)).
\end{equation*}
The family and parameter domain are fixed on each edge, while the two parameter functions may vary with the date and raw market history. Different edges may use different families. Each model computes its conditional uniforms using its own pair copulas, so their joint-law comparison must account for changes in both the pair functions and their conditional inputs.

We use the two oriented conditional transforms
\begin{equation*}
 h_e^{\rightarrow}(u\mid v;\theta_e)=\int_0^u c_e(z,v;\theta_e)\,dz,
 \qquad
 h_e^{\leftarrow}(u\mid v;\theta_e)=\int_0^u c_e(v,z;\theta_e)\,dz.
\end{equation*}
In both orientations, the first argument is the cumulative-probability input of the conditioned variable. Set $\mathcal F_e=\{\log c_e,h_e^{\rightarrow},h_e^{\leftarrow}\}$. For $0<\tau<1/4$, define the local derivative bound
\begin{equation*}
 L_e(\tau)
 =\sup_{\substack{\theta_e\in\Theta_e\\u,v\in[\tau,1-\tau]}}
       \max_{f\in\mathcal F_e}
       \max\{\|\nabla_{\theta_e}f\|_2,|\partial_u f|,|\partial_v f|\},
\end{equation*}
where all derivatives are evaluated at $(u,v;\theta_e)$.

\begin{assumption}
\label{ass:vine_regular}
The pair-copula families satisfy the following conditions.
\begin{enumerate}
 \item Each $\Theta_e$ is nonempty, compact and convex. For every $\theta_e\in\Theta_e$, $c_e(\cdot,\cdot;\theta_e)$ is a valid copula density, strictly positive on $(0,1)^2$, with a version satisfying
 \begin{equation*}
  \int_0^1c_e(z,v;\theta_e)\,dz=1,
  \qquad
  \int_0^1c_e(u,z;\theta_e)\,dz=1,
  \qquad u,v\in(0,1).
 \end{equation*}
 \item The functions in $\mathcal F_e$ are continuously differentiable in the interior inputs and along parameter segments in $\Theta_e$, with the required one-sided derivatives at parameter endpoints. The functions and their derivatives are jointly continuous in the interior inputs and parameters. The auxiliary and fitted parameter functions are measurable in $H$.
 \item There is a finite constant $A\ge1$, depending only on the selected families, parameter domains and parameterisations, such that
 \begin{equation*}
  L_e(\tau)\le\frac A\tau,
  \qquad e\in\mathcal E,\quad 0<\tau<\frac14.
 \end{equation*}
\end{enumerate}
\end{assumption}

We measure the approximation error within the parameter space of each edge by
\begin{equation*}
 \varepsilon_{e,t}(H)
 =\|\theta^\circ_{e,t}(H)-\widehat\theta_{e,t}(H)\|_2.
\end{equation*}

\begin{assumption}
\label{ass:vine_approximation_budget}
For the fitted margins and selected vine, there exists a choice of measurable auxiliary kernels in Equation~\eqref{eq:vine_reference_model} satisfying Assumption~\ref{ass:vine_regular}, and constants $0\le\varepsilon\le1$ and $\zeta\ge0$ such that
\begin{equation*}
 \varepsilon_{e,t}(H)\le\varepsilon,
 \qquad
 D_{\mathrm{KL}}\bigl(q_t(\cdot\mid H)\Vert q_t^\circ(\cdot\mid H)\bigr)
 \le\frac{\zeta^2}{2}
\end{equation*}
for every edge $e$, date $t$ and admissible history $H$. Both bounds use the same auxiliary model. The constants $\varepsilon$ and $\zeta$ are independent of the edge, date and history.
\end{assumption}

The budget $\zeta$ includes errors in the fitted margins and the expressive restrictions of the chosen pair families, vine structure and simplifying assumption. The local derivative condition in Assumption~\ref{ass:vine_regular} covers a wide range of copula families, permitting pair densities to vanish or become unbounded near the boundary, as illustrated by the following families.

\begin{example}
\label{ex:regular_pair_copulas}
Assumption~\ref{ass:vine_regular} holds for the following families and parameter domains, with both models using the same family on each edge and parameters measurable in $H$. The densities below are defined for $(u,v)\in(0,1)^2$.
\begin{enumerate}
 \item Student-$t$: $|\rho|\le\rho_*<1$ and $0<\nu_-\le\nu\le\nu_+<\infty$, with density
 \begin{equation*}
  c_{\rho,\nu}(u,v)
  =\frac{\Gamma((\nu+2)/2)\Gamma(\nu/2)}
         {\sqrt{1-\rho^2}\,\Gamma((\nu+1)/2)^2}
   \frac{\bigl[(1+x^2/\nu)(1+y^2/\nu)\bigr]^{(\nu+1)/2}}
        {\left[1+\dfrac{x^2-2\rho xy+y^2}{\nu(1-\rho^2)}\right]^{(\nu+2)/2}},
 \end{equation*}
 where $x=T_\nu^{-1}(u)$, $y=T_\nu^{-1}(v)$ and $T_\nu$ is the standard univariate Student-$t$ distribution function.
 \item Gaussian: $|\rho|\le\rho_*<1$, with density
 \begin{equation*}
  c_\rho(u,v)=\frac1{\sqrt{1-\rho^2}}
  \exp\!\left\{\frac{2\rho zw-\rho^2(z^2+w^2)}{2(1-\rho^2)}\right\},
 \end{equation*}
 where $z=\Phi^{-1}(u)$, $w=\Phi^{-1}(v)$ and $\Phi$ is the standard normal distribution function.
 \item Clayton: $0\le\theta\le\theta_+<\infty$, with density
 \begin{equation*}
  c_\theta(u,v)=
  \begin{cases}
   (1+\theta)(uv)^{-1-\theta}
   (u^{-\theta}+v^{-\theta}-1)^{-2-1/\theta},&\theta>0,\\
   1,&\theta=0.
  \end{cases}
 \end{equation*}
 The value at $\theta=0$ is the continuous independence extension.
\end{enumerate}
\end{example}

\begin{proof}
See \hyperref[app:vine_regularity_examples]{Appendix~\ref*{app:dependence_results_proofs}}.
\end{proof}

We next construct the error bound along the complete vine. Fix a date and history, suppress $(t,H)$ temporarily, and choose a common evaluation recursion for the auxiliary, fitted and parameter-interpolated models. At level $k$, every density factor has inputs whose conditioning sets have size $k-1$. We evaluate both oriented conditional outputs of each edge in levels $1,\ldots,d-2$. Together with the original uniforms, these outputs contain all recursive inputs to the $d-1$ levels of density factors. When a conditional uniform has multiple recursive representations, we fix the same representation throughout.

For $0<\tau<1/4$, set $r_{i\mid\varnothing}(\tau)=0$. For $e=(i,j;D)\in\mathcal E_k$, $1\le k\le d-2$, define
\begin{equation}
 r_{i\mid D\cup\{j\}}(\tau)
 =L_e(\tau)\bigl[\varepsilon_e+r_{i\mid D}(\tau)+r_{j\mid D}(\tau)\bigr].
 \label{eq:vine_uniform_recursion}
\end{equation}
The reverse orientation interchanges $i$ and $j$. The quantity $r_{i\mid D}(\tau)$ bounds the derivative of that conditional uniform along the segment joining the two parameter vectors whenever its recursive inputs remain in $[\tau,1-\tau]$. Restoring $(t,H)$, define
\begin{align}
 \Delta_t(H;\tau)
 ={}&\sum_{e=(i,j;D)\in\mathcal E}
 L_e(\tau)\Bigl[\varepsilon_{e,t}(H)
       +r_{i\mid D,t}(H;\tau)+r_{j\mid D,t}(H;\tau)\Bigr]\notag\\
 &+\frac4\tau\sum_{k=1}^{d-2}\sum_{e=(i,j;D)\in\mathcal E_k}
 \Bigl[r_{i\mid D\cup\{j\},t}(H;\tau)
       +r_{j\mid D\cup\{i\},t}(H;\tau)\Bigr],\notag\\
 \Delta_t(H)
 ={}&\inf_{0<\tau<1/4}\bigl\{\Delta_t(H;\tau)+8d\sqrt\tau\bigr\}.
 \label{eq:vine_error_budget}
\end{align}
The first sum controls all density factors. The second controls the derivatives of the localisation factors for the two oriented outputs in the first $d-2$ levels and is empty when $d=2$. At each fixed $\tau$, $\Delta_t(H;\tau)$ is a nonnegative linear combination of the parameter errors. The infimum then balances this local contribution against the boundary term. Continuity of the derivatives and compactness of the parameter domains imply continuity of $L_e(\tau)$. Thus the infimum may equivalently be taken over rational $\tau$, which ensures measurability of $\Delta_t(H)$ in $H$.

We use the unnormalised Hellinger distance and the total-variation convention
\begin{equation*}
 h(P,Q)^2=\int(\sqrt p-\sqrt q)^2\,d\mu,
 \qquad
 \|P-Q\|_{\mathrm{TV}}=\sup_B|P(B)-Q(B)|,
\end{equation*}
where $p,q$ are densities with respect to a common dominating measure $\mu$.

\begin{theorem}
\label{thm:vine_density_stability}
Under Assumption~\ref{ass:vine_regular}, the auxiliary and fitted kernels in Equation~\eqref{eq:vine_reference_model} satisfy, at every date and history,
\begin{align}
 h\bigl(q_t^\circ(\cdot\mid H),\widehat q_t(\cdot\mid H)\bigr)
 &\le\frac{\Delta_t(H)}2,\notag\\
 \|q_t^\circ(\cdot\mid H)-\widehat q_t(\cdot\mid H)\|_{\mathrm{TV}}
 &\le\min\left\{1,\frac{\Delta_t(H)}2\right\}.
 \label{eq:vine_density_stability}
\end{align}
If Assumption~\ref{ass:vine_approximation_budget} also holds, then
\begin{align}
 \Delta_t(H;\tau)
 &\le7(2A)^{d-1}\varepsilon\tau^{-(d-1)},
 \qquad 0<\tau<\frac14,\notag\\
 \Delta_t(H)
 &\le 8^dA^{d-1}\varepsilon^{1/(2d-1)}, \qquad 0\le\varepsilon\le1.
 \label{eq:vine_uniform_budget}
\end{align}
For simplicity, we write $C_d=8^dA^{d-1}$ in the rest of the paper.
\end{theorem}

\begin{proof}
See \hyperref[app:vine_density_proof]{Appendix~\ref*{app:dependence_results_proofs}}.
\end{proof}

To include marginal and structural misspecification, we use the relative entropy between the true and auxiliary laws on the return space. Assumption~\ref{ass:vine_approximation_budget} bounds this discrepancy by $\zeta^2/2$. Pinsker's inequality and the Hellinger--entropy inequality convert it into distance bounds, which we combine with Theorem~\ref{thm:vine_density_stability}.

\begin{lemma}
\label{lem:vine_model_error}
Under Assumptions~\ref{ass:vine_regular} and \ref{ass:vine_approximation_budget}, at every date and history,
\begin{align}
 \|q_t(\cdot\mid H)-\widehat q_t(\cdot\mid H)\|_{\mathrm{TV}}
 &\le\min\left\{1,\frac{\zeta+C_d\varepsilon^{1/(2d-1)}}2\right\},\notag\\
 h\bigl(q_t(\cdot\mid H),\widehat q_t(\cdot\mid H)\bigr)
 &\le\frac{\zeta}{\sqrt2}+\frac{C_d\varepsilon^{1/(2d-1)}}2
 \le\frac{\zeta+C_d\varepsilon^{1/(2d-1)}}{\sqrt2}.
 \label{eq:vine_total_law_error}
\end{align}
\end{lemma}

\begin{proof}
See \hyperref[app:vine_model_error_proof]{Appendix~\ref*{app:dependence_results_proofs}}.
\end{proof}

\subsection{Stability of dynamic portfolio values}
\label{subsec:vine_policy_performance}

The market path laws $\mathbb P$ and $\widehat{\mathbb P}$ are obtained by iterating $q_t$ and $\widehat q_t$ from the common initial history. Their investment criteria $J$ and $\widehat J$ use their respective conditional CVaR charges. Let $\Pi$ be a nonempty common class of admissible policies in Definition~\ref{def:admissible_policy}, and write $V^\Pi=\sup_{\pi\in\Pi}J(\pi;s_0)$ and $\widehat V^\Pi=\sup_{\pi\in\Pi}\widehat J(\pi;s_0)$. A fixed policy generates the same wealth and holdings on the same realised return path under either model.

We require the common policy class to satisfy the terminal-utility moment condition
\begin{equation}
 \sup_{\pi\in\Pi}\max\left\{
 \mathbb E\left[|U_{\gamma_U}(X_T^\pi)|^2\right],
 \widehat{\mathbb E}\left[|U_{\gamma_U}(X_T^\pi)|^2\right]
 \right\}\leq M_T<\infty.
 \label{eq:vine_terminal_moment}
\end{equation}
For $\gamma_U\ne1$, Equation~\eqref{eq:vine_terminal_moment} bounds the expectation of $(X_T^\pi)^{2(1-\gamma_U)}$ under each model by $(1-\gamma_U)^2M_T$. For $\gamma_U=1$, it controls $|\log X_T^\pi|^2$. For every fixed $\gamma_U>1$, Lemma~\ref{lem:reachable_regularity} gives $X_T^\pi\geq\underline k^{\,T}$, so Equation~\eqref{eq:vine_terminal_moment} holds under both models, uniformly over admissible policies, with $M_T=(\gamma_U-1)^{-2}\underline k^{-2T(\gamma_U-1)}$. Combining this moment budget with the conditional CVaR bounds from Assumption~\ref{ass:market}, we define the horizon coefficient
\begin{equation}
 \mathcal K_T
 =\sqrt{2TM_T}
 +\lambda(B-\underline k)
       \left(\frac{T^{3/2}}{3}+\frac{T}{2\alpha p}\right),
 \label{eq:vine_performance_bound}
\end{equation}
where $p=1-\alpha$.

\begin{theorem}
\label{thm:vine_policy_performance}
Suppose Assumption~\ref{ass:market} holds for the true and fitted models, Assumptions~\ref{ass:vine_regular} and \ref{ass:vine_approximation_budget} hold, and the nonempty common policy class $\Pi$ satisfies Equation~\eqref{eq:vine_terminal_moment}. Then
\begin{align}
 \sup_{\pi\in\Pi}|J(\pi;s_0)-\widehat J(\pi;s_0)|
 &\le\mathcal K_T(\zeta+C_d\varepsilon^{1/(2d-1)}),\notag\\
 |V^\Pi-\widehat V^\Pi|&\le\mathcal K_T(\zeta+C_d\varepsilon^{1/(2d-1)}).
 \label{eq:vine_value_stability}
\end{align}
\end{theorem}

\begin{proof}
See \hyperref[app:vine_performance_proof]{Appendix~\ref*{app:dependence_results_proofs}}.
\end{proof}

\begin{remark}
If $M_T\leq C(1+T)^\nu$ for constants $C<\infty$ and $\nu\geq0$ independent of the horizon and policy, then, with the remaining model constants fixed, Equation~\eqref{eq:vine_performance_bound} gives
\begin{equation*}
 \mathcal K_T=O\bigl(T^{(\nu+1)/2}+T^{3/2}\bigr).
\end{equation*}
In particular, $M_T=O(T^2)$ yields $\mathcal K_T=O(T^{3/2})$. For logarithmic utility, Assumption~\ref{ass:market} directly gives, uniformly over all admissible policies,
\begin{equation}
 \max\left\{\mathbb E|\log X_T^\pi|^2,
 \widehat{\mathbb E}|\log X_T^\pi|^2\right\}
 \leq T^2\left[(\max\{0,-\log\underline k\})^2+\frac{4B}{e^2}\right].
 \label{eq:vine_log_moment}
\end{equation}
Thus logarithmic utility admits a budget $M_T=O(T^2)$ when $B$ and $\underline k$ are independent of $T$. We prove Equation~\eqref{eq:vine_log_moment} at the end of the proof of Theorem~\ref{thm:vine_policy_performance}.
\end{remark}

\begin{remark}
The date-specific budgets $\delta_t=\sup_H\Delta_t(H)$ give a finer comparison. Let $\mathbb P_{[t]}$ and $\widehat{\mathbb P}_{[t]}$ denote the laws of the first $t$ returns under the true and fitted models, starting from the common initial history. We accumulate the true-to-auxiliary errors by the KL chain rule and the auxiliary-to-fitted errors by the conditional Hellinger identity. The resulting path bounds are
\begin{align*}
 \|\mathbb P_{[t]}-\widehat{\mathbb P}_{[t]}\|_{\mathrm{TV}}
 &\le\frac12\left[\zeta\sqrt t+\left(\sum_{r<t}\delta_r^2\right)^{1/2}\right],\\
 h(\mathbb P_{[t]},\widehat{\mathbb P}_{[t]})
 &\le\zeta\sqrt{t/2}+\frac12\left(\sum_{r<t}\delta_r^2\right)^{1/2}.
\end{align*}
At a fixed history, the conditional CVaR comparison uses the total-variation budget $(\zeta+\delta_t)/2$. Substituting $\delta_t\le C_d\varepsilon^{1/(2d-1)}$ recovers the uniform estimate in Theorem~\ref{thm:vine_policy_performance}.
\end{remark}

In Theorem~\ref{thm:vine_policy_performance}, the marginal and structural term is $\mathcal K_T\zeta$, and the pair-copula parameter term is $\mathcal K_TC_d\varepsilon^{1/(2d-1)}$. Errors in early conditional laws also affect the distribution of later wealth and holdings, while the current CVaR estimate amplifies conditional-law error by $(B-\underline k)/(\alpha p)$. Local propagation through the conditional inputs and control of the boundary probability produce the exponent $1/(2d-1)$ and the constant $C_d=8^dA^{d-1}$. For fixed dimension, parameter domains and horizon, this contribution tends to zero with the parameter error. If $q_t^\circ=q_t$ is admissible, we may take $\zeta=0$ and recover the common-margin, common-structure comparison.

\subsection{Performance of policies selected under the fitted model}
\label{subsec:vine_policy_transfer}

For deployment, we evaluate the policy selected under $\widehat{\mathbb P}$ using the true-market criterion $J$, including the conditional CVaR charges computed under the true-market kernel $q_t$. The same uniform comparison controls this loss.

\begin{theorem}
\label{thm:vine_policy_transfer}
Under the assumptions of Theorem~\ref{thm:vine_policy_performance}, suppose $\widehat\pi\in\Pi$ satisfies $\widehat V^\Pi-\widehat J(\widehat\pi;s_0)\le\eta$ for some $\eta\ge0$. Then
\begin{equation}
 0\le V^\Pi-J(\widehat\pi;s_0)\le2\mathcal K_T(\zeta+C_d\varepsilon^{1/(2d-1)})+\eta.
 \label{eq:vine_transfer_bound}
\end{equation}
\end{theorem}

\begin{proof}
See \hyperref[app:vine_transfer_proof]{Appendix~\ref*{app:dependence_results_proofs}}.
\end{proof}

For a simulated training criterion, we can separate approximation of the fitted objective from its numerical optimisation.

\begin{corollary}
\label{cor:vine_learning}
Under the assumptions of Theorem~\ref{thm:vine_policy_performance}, suppose a simulated criterion $J_{\mathrm{sim}}$ and a returned policy $\widehat\pi\in\Pi$ satisfy, on an event under consideration,
\begin{equation}
 \sup_{\pi\in\Pi}|J_{\mathrm{sim}}(\pi)-\widehat J(\pi;s_0)|\le E_{\mathrm{sim}},
 \qquad
 \sup_{\pi\in\Pi}J_{\mathrm{sim}}(\pi)-J_{\mathrm{sim}}(\widehat\pi)
 \le\eta_{\mathrm{opt}}.
 \label{eq:vine_learning_conditions}
\end{equation}
Then, on the same event,
\begin{equation}
 V^\Pi-J(\widehat\pi;s_0)
 \le2\mathcal K_T\zeta+2\mathcal K_TC_d\varepsilon^{1/(2d-1)}
      +2E_{\mathrm{sim}}+\eta_{\mathrm{opt}}.
 \label{eq:vine_learning_bound}
\end{equation}
\end{corollary}

\begin{proof}
See \hyperref[app:vine_learning_proof]{Appendix~\ref*{app:dependence_results_proofs}}.
\end{proof}

Theorem~\ref{thm:vine_policy_transfer} controls the performance loss when transferring a fitted-model policy to the true market. Corollary~\ref{cor:vine_learning} incorporates simulation and numerical optimisation errors under Equation~\eqref{eq:vine_learning_conditions}. Quantifying $E_{\mathrm{sim}}$ and $\eta_{\mathrm{opt}}$ for the learning algorithm remains a direction for future study.

Equation~\eqref{eq:vine_learning_bound} separates four contributions to the true-market performance loss: marginal and structural approximation, pair-copula parameter approximation, simulation, and numerical optimisation. The detailed vine recursion determines the second contribution and its uniform coefficient $C_d$. The first is controlled by the abstract approximation budget $\zeta$ in Assumption~\ref{ass:vine_approximation_budget}, allowing the true market to lie outside the selected simplified-vine class.

\section{Numerical implementation}
\label{sec:setup_implementation}

We implement the dependence-based control approach of Section~\ref{sec:problem_formulation} through a fitted return generator, recurrent policy, and fixed training and evaluation rules. Section~\ref{sec:dependence_robustness} studies an idealised investment problem and provides theoretical justification. The numerical study then verifies the empirical performance of the proposed approach in real data.

\subsection{From conditional returns to portfolio decisions}
\label{subsec:algorithmic_map}

\begin{figure}[ht]
\centering
\includegraphics[width=420.33412pt,height=195.12349pt]{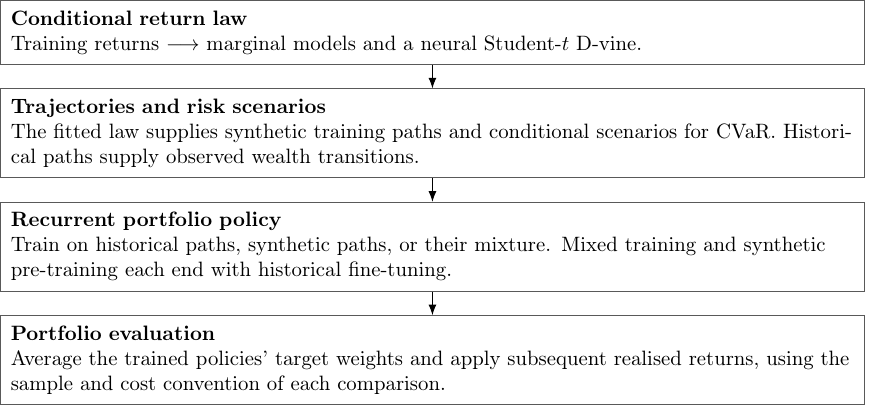}
\caption{Conditional return modelling, policy learning, and portfolio evaluation. Estimation and policy fitting use the training sample. During evaluation, each decision uses the available history, followed by the realised holding-period return.}
\label{fig:complete_algorithm_map}
\end{figure}

We use the fitted return law to generate training trajectories and conditional risk scenarios. The recurrent policy maps the observed history to portfolio weights, which determine wealth and risk-adjusted rewards. Figure~\ref{fig:complete_algorithm_map} summarises these dependencies.

\subsection{Wealth dynamics and training objective}
\label{subsec:numerical_accounting}

We use $d=7$ risky-asset target weights $w_t$, residual cash weight $w_t^0=1-\mathbf1^\top w_t$, holding-period gross returns $G_{t+1}$, and cash gross return $R^f_{t+1}$. The pre-cost portfolio multiplier is $\Gamma_{t+1}(w_t)=w_t^0R^f_{t+1}+w_t^\top G_{t+1}$. The portfolio constraints are
\begin{equation}
\mathcal A=\left\{w\in\mathbb R^d:
\mathbf1^\top w=1,\ \lVert w\rVert_1\leq1.5,
-0.20\leq w_i\leq0.60\right\}.
\label{eq:action_space}
\end{equation}
The policy and optimised benchmarks use the feasible set in Equation~\eqref{eq:action_space}, so the target cash weight is zero.

For a positive pre-cost multiplier, realised holding-period returns produce the pre-trade weights $\widetilde w_{i,t}=w_{i,t-1}G_{i,t}/\Gamma_t(w_{t-1})$. The training environment uses drift-aware full-$\ell_1$ turnover $\operatorname{TO}_t(w)=\lVert w-\widetilde w_t\rVert_1$. The benchmark, training-scheme, and policy-input evaluations in Tables~\ref{tab:outofsample}, \ref{tab:curriculum_comparison}, and~\ref{tab:information_comparison} use target-to-target turnover $\operatorname{TO}_t(w)=\lVert w-w_{t-1}\rVert_1$ and a fixed financing year fraction $\Delta_t=1/12$. With transaction-cost rate $\kappa$, annual short- and cash-borrow rates $c_s,c_c$, and year fraction $\Delta_t$, the corresponding cost multiplier is
\begin{equation*}
m_t(w)=\exp\!\left[-\kappa\operatorname{TO}_t(w)
-\Delta_t\left\{c_s\sum_i(-w_i)_+ +c_c(-w^0)_+\right\}\right].
\end{equation*}
The evaluator updates wealth by
\begin{equation}
V_{t+1}=V_t\max\!\left\{m_t(w_t)\Gamma_{t+1}(w_t),\epsilon_V\right\},
\qquad\epsilon_V=10^{-12}.
\label{eq:cost_adjusted_wealth}
\end{equation}
The floor prevents undefined utility evaluations.

For $M=512$ conditional scenarios from the fitted generator, we compute losses $L_{t+1}^{(m)}(w)=1-m_t(w)\Gamma_{t+1}^{(m)}(w)$ and the empirical tail mean
\begin{equation}
\widehat{\operatorname{CVaR}}_{\alpha,t}(w)
=\frac{1}{k_\alpha}\sum_{m\in\mathcal I_{\alpha,t}(w)}L_{t+1}^{(m)}(w),
\qquad k_\alpha=\lceil(1-\alpha)M\rceil,
\qquad\alpha=0.95.
\label{eq:scenario_cvar}
\end{equation}
Here $\mathcal I_{\alpha,t}(w)$ selects the $k_\alpha=26$ largest losses.\footnote{Exact empirical CVaR at $M=512$ and $\alpha=0.95$ averages the 25 largest losses and $0.6$ of the 26th, divided by $25.6$.}

With $U_{\gamma_U}(x)=(x^{1-\gamma_U}-1)/(1-\gamma_U)$ and $\gamma_U=2$, the training reward is
\begin{equation}
r_t=U_{\gamma_U}\!\left(\frac{V_{t+1}}{V_0}\right)
-U_{\gamma_U}\!\left(\frac{V_t}{V_0}\right)
-\lambda\widehat{\operatorname{CVaR}}_{\alpha,t}(w_t),
\qquad\lambda=0.05.
\label{eq:reward_function}
\end{equation}
Thus the undiscounted rewards in Equation~\eqref{eq:reward_function} sum to terminal wealth utility minus the sequence of scenario-risk charges, since $U_{\gamma_U}(1)=0$. For $K$ presented episodes drawn from an episode distribution $\mathbb Q$ and policy weights $\phi$, the corresponding empirical criterion is
\begin{equation}
J_{\mathrm{emp},K}(\phi;\mathbb Q)=\frac1K\sum_{k=1}^K\left[
U_{\gamma_U}\!\left(\frac{V_{T,k}^{\pi_\phi}}{V_0}\right)
-\lambda\sum_{t=0}^{T-1}\widehat{\operatorname{CVaR}}_{\alpha,t,k}(\pi_\phi)
\right].
\label{eq:empirical_control_problem}
\end{equation}
TD3 uses Equation~\eqref{eq:empirical_control_problem} as its economic objective through replay and critic-based actor updates, with the actor regularisers specified in Section~\ref{subsec:learning_algorithm}. During historical fine-tuning, historical observations determine the realised wealth transitions while the fitted vine determines the scenario-risk charge.

\subsection{Conditional return modelling and scenario generation}
\label{subsec:conditional_law_estimation}
\label{subsec:3_Dvine_copula_model}

Marginal models describe conditional location and scale, and a dynamic D-vine represents the conditional dependence of the resulting innovations. Every fit uses observations dated no later than the relevant decision date.

\subsubsection{Marginal return models}
\label{subsubsec:3_marginal_models}

For asset $i$, the candidate conditional mean is either constant or AR(1),
\begin{equation*}
r_{i,t}=\mu_i+\phi_i r_{i,t-1}+\sigma_{i,t}z_{i,t},
\qquad \phi_i=0\ \text{for the constant-mean candidate},
\end{equation*}
and the volatility grid comprises GARCH$(1,1)$ \autocite{bollerslevGeneralizedAutoregressiveConditional1986}, GJR--GARCH \autocite{glostenRelationExpectedValue1993}, and eGARCH$(1,1)$ \autocite{nelsonConditionalHeteroskedasticityAsset1991} specifications. Innovations follow either a Student-$t$ distribution or the skewed Student-$t$ construction of \textcite{fernandezBayesianModelingFat1998}. The symmetric GARCH candidate is
\begin{equation*}
\sigma_{i,t}^2=\omega_i+\alpha_i\varepsilon_{i,t-1}^2+\beta_i\sigma_{i,t-1}^2,
\qquad \omega_i>0,\quad \alpha_i,\beta_i\geq0,\quad \alpha_i+\beta_i<1.
\end{equation*}
We retain converged candidates whose Ljung--Box tests on $z_{i,t}$ and $z_{i,t}^2$ at lags 5, 10, and 20 all have $p$-values of at least 0.01 \autocite{ljungMeasureLackFit1978}. We select among them by the Bayesian information criterion \autocite{schwarzEstimatingDimensionModel1978}. If no parametric candidate passes these checks, we test a two-component EWMA grid under the same criteria and stop if none passes.

The fitted conditional distribution $F_{i,t}$ gives the probability integral transform
\begin{equation}
u_{i,t}=F_{i,t}(r_{i,t}\mid\mathcal F_{t-1}),
\qquad u_{i,t}\in(0,1),
\label{eq:pit}
\end{equation}
which separates dependence from marginal scale. Parameters and filtering states are fixed at the training cutoff for episode and evaluation generation.

\subsubsection{Neural dynamic D-vine}
\label{subsubsec:3_dynamic_vine_with_NN_param}

For a fixed order $1-\cdots-d$, the pair-copula construction factorises the copula density as \autocite{bedfordProbabilityDensityDecomposition2001a,bedfordVinesNewGraphical2002a}
\begin{equation*}
c_t(u_{1:d})=
\prod_{\ell=1}^{d-1}\prod_{k=1}^{d-\ell}
c_{k,k+\ell\mid k+1:k+\ell-1,t}
\!\left(u_{k\mid k+1:k+\ell-1,t},
u_{k+\ell\mid k+1:k+\ell-1,t}\right),
\end{equation*}
where higher-tree arguments follow recursively from pair-copula $h$-functions \autocite{aasPaircopulaConstructionsMultiple2009}. Training-prefix Kendall dependence selects the order once. For $d=7$, all 21 edges use Student-$t$ copulas: edge degrees of freedom $\nu_e$ are fixed after training-prefix estimation, while correlations vary through time.

For edge $e$, the information vector $x_{e,t}$ contains lagged conditional pair inputs, their interaction, the previous correlation, and the lagged copula score. A two-hidden-layer network produces
\begin{equation}
\rho_{e,t}=\tanh\!\left(\operatorname{atanh}(\bar\rho_e)
+a_e\tanh f_{\psi_e}(x_{e,t})\right),
\label{eq:nn_parameter}
\end{equation}
where $\bar\rho_e$ is the static training-prefix estimate and $a_e$ bounds its dynamic departure. The lagged score supplies likelihood-based information, as in score-driven models \autocite{crealGENERALIZEDAUTOREGRESSIVESCORE2013}. Parameters minimise the Student-$t$ copula loss,
\begin{equation}
\widehat\psi_e\in\arg\min_{\psi_e}
-\sum_{t\in\mathcal T_{\rm train}}
\log c_{t,\nu_e}\!\left(u_{e,1,t},u_{e,2,t};\rho_{e,t}(\psi_e)\right),
\label{eq:nn_objective}
\end{equation}
with chronological validation and early stopping.

After likelihood fitting, a training-sample calibration rescales the edge correlations:
\begin{align}
c^\star&=\arg\min_{c>0}
\|\widehat m_{\rm sim}(c)-\widehat m_{\rm train}\|_{W}^{2},\nonumber\\
\widetilde\rho_{e,t}&=\tanh\!\left[c^\star\operatorname{atanh}\{\rho_{e,t}(\widehat\psi_e)\}\right].
\label{eq:dependence_scale_calibration}
\end{align}
The calibrated copula is $C_t^{\rm NN}=C_{\mathcal O^\star}(\{\widetilde\rho_{e,t}\}_{e=1}^{21},\{\nu_e\}_{e=1}^{21})$, where $\mathcal O^\star$ is the estimated asset order.

\subsubsection{Episode and scenario construction}
\label{subsubsec:3_synthetic_data_generation}

Daily filtered innovations are aggregated to monthly holding-period uniforms. A univariate Student-$t$ Markov copula per asset supplies serial persistence. Conditional contemporaneous innovations are then sampled through inverse Rosenblatt recursion from the neural D-vine:
\begin{equation}
U_{i,t+1}=H^{-1}_{i\mid 1:i-1,t}
\left(E_{i,t+1}\mid U_{1:i-1,t+1};\widehat\Theta_t^{\rm vine}\right),
\qquad E_{i,t+1}\stackrel{\mathrm{iid}}{\sim}\mathcal U(0,1).
\label{eq:dvine_sample}
\end{equation}
Monthly log returns are recovered by the fixed empirical marginal quantiles,
\begin{equation}
\widetilde r_{i,t+1}=\widehat F^{-1}_{i,\mathrm{month}}(U_{i,t+1}),
\qquad \widetilde G_{i,t+1}=\exp(\widetilde r_{i,t+1}).
\label{eq:inverse_returns}
\end{equation}

Each synthetic episode has a 30-month burn-in and 24 decisions. At each synthetic decision, one joint draw defines the transition and $M=512$ independent draws estimate the risk charge in Equation~\eqref{eq:scenario_cvar}. The main model uses 100 distinct synthetic paths within the mixed presentation budget specified in Section~\ref{sec:causal_component_analysis}. Appendix~B reports generative-fidelity diagnostics for these paths and pre-training diagnostics for the same ten compressed-state policies. We draw episode starting states from the training sample. Historical fine-tuning uses the wealth transitions and scenario-risk charges defined in Section~\ref{subsec:numerical_accounting}.

Algorithm~\ref{alg:neural_vine_generator} gives the estimation and sampling steps. \hyperref[subsec:7_generative_fidelity_training_dynamics]{Appendix B} examines marginal, dependence, tail, and temporal fidelity.

\begin{algorithm}[htb]
\caption{Neural D-vine scenario law and episode construction}
\label{alg:neural_vine_generator}
\footnotesize
\begin{algorithmic}[1]
\REQUIRE Training returns $D_{\rm train}$, episode count $N_{\rm synth}$, burn-in $L$, horizon $T$, and risk-scenario count $M$
\ENSURE Fitted conditional-law estimator $\widehat F_t$ and synthetic trajectories $\mathcal D_{\rm synth}$
\FOR{$i=1,\ldots,d$}
  \STATE Select the marginal model and compute the probability integral transforms in Equation~\eqref{eq:pit}.
\ENDFOR
\STATE Select the asset order $\mathcal O^\star$ from training-sample Kendall dependence and estimate the Student-$t$ pair parameters $\{(\bar\rho_e,\nu_e)\}_{e=1}^{21}$.
\FOR{$e=1,\ldots,21$}
  \STATE Fit the correlation network in Equation~\eqref{eq:nn_parameter} by minimising Equation~\eqref{eq:nn_objective}, with chronological validation.
\ENDFOR
\STATE Calibrate the dependence scale and construct $C_t^{\rm NN}$ by Equation~\eqref{eq:dependence_scale_calibration}.
\FOR{$n=1,\ldots,N_{\rm synth}$}
  \STATE Sample a causal snapshot path $(C_{n,t}^{\rm NN})_{t=-L+1}^{T}$.
  \FOR{$t=-L+1,\ldots,T$}
    \STATE Draw $\mathbf U_{n,t}^{(m)}\sim C_{n,t}^{\rm NN}$, $m=0,\ldots,M$, by Equation~\eqref{eq:dvine_sample} and apply the fitted serial transitions.
    \STATE Recover gross returns using Equation~\eqref{eq:inverse_returns}.
  \ENDFOR
  \STATE Form the episode from its initial history, transition draws, and risk scenarios.
\ENDFOR
\STATE Apply the fixed marginal, correlation, tail, and temporal diagnostic criteria.
\RETURN $(\widehat F_t,\mathcal D_{\rm synth})$
\end{algorithmic}
\end{algorithm}

\subsection{Recurrent TD3 implementation}
\label{subsec:3_lstm_td3}
\label{subsec:learning_algorithm}
\label{subsec:observation_implementation}
\label{subsec:3_two_stage_training}

The actor observes a 30-month history of normalised wealth, seven recent returns, seven rolling volatilities, seven drifted pre-trade weights, and their gross and net exposures. All information variants use the same network architecture and an 88-dimensional input. The compressed-state policy uses these 24 coordinates and sets the remaining 64 to zero. The information comparisons activate the scalar scenario-CVaR observation, the 63 vine features, or both. Each of the 21 vine edges contributes its correlation, implied Kendall's $\tau$, and symmetric tail-dependence coefficient. The vine supplies simulated training paths and the reward's scenario-risk charge.

The continuous-action problem in Equation~\eqref{eq:empirical_control_problem} is approximated by recurrent Twin Delayed Deep Deterministic Policy Gradient (TD3) \autocite{fujimotoAddressingFunctionApproximation2018}. A layer-normalised long short-term memory (LSTM) network encodes $H_t^{(30)}\in\mathbb R^{30\times88}$ \autocite{hochreiterLongShortTermMemory1997,baLayerNormalization2016}. We write $H_t=H_t^{(30)}$ below.

The actor encoder maps the normalised sequence into its final hidden representation,
\begin{equation*}
h_t^\mu=\operatorname{LN}\!\left(\operatorname{LSTM}_{\phi}
(\operatorname{LN}(H_t^{(30)}))_{30}\right)\in\mathbb R^{64},
\qquad (z_t,z_t^{\rm lev})=f_{\phi}(h_t^\mu),
\end{equation*}
where $f_\phi$ is the output head. We denote the complete raw-action map, comprising the encoder and output head, by $a_\phi(H_t)=(z_t,z_t^{\rm lev})$.

The sigmoid coefficient $g_t=\operatorname{sigmoid}(z_t^{\rm lev})$ sets the short and long budgets $B_t^-=0.25g_t$ and $B_t^+=1+B_t^-$. The two lowest scores form the possible short support. Temperature-scaled softmax vectors on the disjoint supports are projected onto simplices capped at $0.20/B_t^-$ and $0.60/B_t^+$, respectively. The resulting policy is
\begin{equation}
w_t=\mu_\phi(H_t)=\mathcal P_{\mathcal A}(z_t,z_t^{\rm lev})
=B_t^+p_t^+(z_t)-B_t^-p_t^-(z_t).
\label{eq:constrained_action_map}
\end{equation}
Here $\mathcal P_{\mathcal A}$ denotes the rank-based action map, whose range is a proper subset of $\mathcal A$. The emitted target has net exposure one, gross exposure $1+0.5g_t\leq1.5$, and the position bounds in Equation~\eqref{eq:action_space}. Exploration and target-policy noise are applied before this map. Each critic uses its own history encoder and parameters to estimate $Q_{\vartheta_j}(H_t,w_t)$ from the encoded history and constrained action.

For a replay transition $(H_t,w_t,r_t,H_{t+1},d_t)$, clipped Gaussian smoothing is applied in raw action space and passed through the action map in Equation~\eqref{eq:constrained_action_map}:
\begin{align*}
w'_{t+1}&=\mathcal P_{\mathcal A}
\left(a_{\phi^-}(H_{t+1})+\operatorname{clip}(\epsilon,-c,c)\right),
\qquad \epsilon\sim\mathcal N(0,\sigma_{\rm targ}^2I),\\
y_t&=r_t+\gamma(1-d_t)
\min_{j\in\{1,2\}}Q_{\vartheta_j^-}(H_{t+1},w'_{t+1}).
\end{align*}
The critics minimise
\begin{equation*}
\mathcal L_Q(\vartheta_j)=
\mathbb E_{\mathcal B}\left[(Q_{\vartheta_j}(H_t,w_t)-y_t)^2\right],
\end{equation*}
and every second critic update the actor maximises
\begin{equation}
\mathcal J_\mu(\phi)=
\mathbb E_{\mathcal B}\left[Q_{\vartheta_1}(H_t,\mu_\phi(H_t))\right]
+\beta_H\mathbb E_{\mathcal B}[\mathcal H(p_t^+)]
-\beta_L\mathbb E_{\mathcal B}[(g_t-g_0)^2].
\label{eq:actor_objective}
\end{equation}
The actor update follows the deterministic policy gradient \autocite{silverDeterministicPolicyGradient2014}, and twin critics, delayed updates, and policy smoothing follow TD3 \autocite{fujimotoAddressingFunctionApproximation2018}. In Equation~\eqref{eq:actor_objective}, $\beta_H$ weights the entropy term $\mathcal H(p_t^+)$, and $\beta_L$ weights the squared deviation of $g_t$ from the reference coefficient $g_0$. Targets satisfy $\xi^-\leftarrow\tau\xi+(1-\tau)\xi^-$, and decaying Gaussian exploration noise is applied before the same action map. The discount factor $\gamma=1$ preserves the undiscounted empirical objective defined in Section~\ref{subsec:numerical_accounting}.

\begin{figure}[tbp]
\centering
\includegraphics[width=477.66061pt,height=173.20395pt]{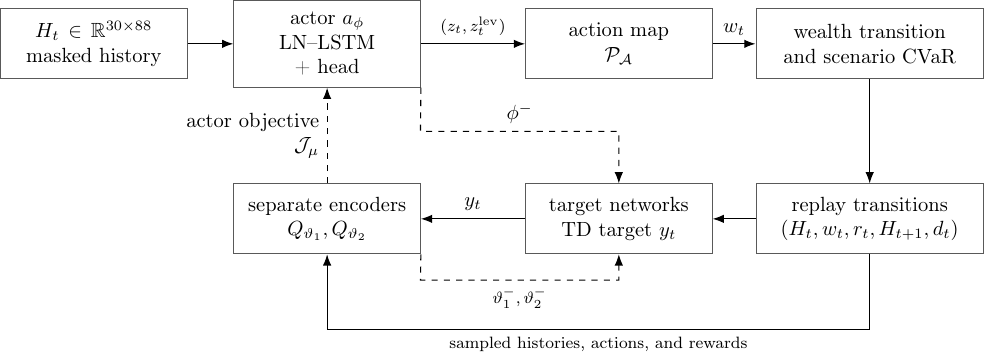}
\caption{Recurrent TD3. The actor maps the masked history to portfolio weights, and realised transitions and scenario-risk charges supply the replay rewards. The two critics use separate history encoders. Dashed arrows indicate parameter updates.}
\label{fig:lstm_td3_computation_graph}
\end{figure}

Figure~\ref{fig:lstm_td3_computation_graph} shows how portfolio transitions supply the replay data for the actor and critic updates.

Training uses historical trajectories, simulated trajectories, or their mixture, as specified in Section~\ref{sec:causal_component_analysis}. For the mixed scheme, a separate fit on 37 early historical trajectories diagnoses one-pass historical adaptation on a later trajectory with disjoint targets, with per-seed metrics retained in the run records. Final historical fine-tuning starts from the pre-trained parameters with a fresh replay buffer and reset target networks, and makes one pass through all 61 historical training paths. Appendix~A lists the numerical settings.

At each evaluation date, each independently trained policy $s\in\mathcal S$ produces target weights $w_t^{(s)}$ from its own history. We form the ensemble target $\bar w_t=|\mathcal S|^{-1}\sum_{s\in\mathcal S}w_t^{(s)}$ before applying realised returns and transaction costs. Policy parameters remain fixed throughout evaluation.

\subsection{Benchmarks and information controls}
\label{subsec:3_benchmark_models}

Within each comparison, benchmark policies share the dates, asset order, realised returns, cost convention specified in Section~\ref{subsec:numerical_accounting}, and feasible set $\mathcal A$. Decisions at $t$ may use $\mathcal F_t$ but no later observations.

\label{subsubsec:3_classical_models}
\label{subsubsec:3_vine_based_models}

Equal weight is the fixed policy $w_t=\mathbf1/d$. The shrinkage mean--variance policy solves \autocite{markowitzPortfolioSelection1952b}
\begin{equation}
w_t^*\in\arg\min_{w\in\mathcal A}
\left\{w^\top\widehat\Sigma_t w-\lambda_{\rm MV}w^\top\widehat\mu_t\right\},
\label{eq:markowitz_constrained}
\end{equation}
using oracle-approximating covariance shrinkage \autocite{chenShrinkageAlgorithmsMMSE2010} and James--Stein-type mean contraction towards zero \autocite{jamesEstimationQuadraticLoss1992}. DCC--GARCH combines conditional univariate scales with recursively updated correlations \autocite{engleDynamicConditionalCorrelation2002a}.

Static, rolling, and neural D-vine benchmarks separate dependence estimation from sequential control. They respectively reuse the training-prefix vine, refit over the available window, or apply Section~\ref{subsubsec:3_dynamic_vine_with_NN_param} without RL. Each vine benchmark draws $N_{\rm bench}=2{,}048$ one-period scenarios from its fitted conditional law and solves
\begin{equation}
w_t^*\in\arg\max_{w\in\mathcal A}
\frac1{N_{\rm bench}}\sum_{m=1}^{N_{\rm bench}}U_\gamma
\!\left(1-c_t(w,\widetilde w_t)+w^\top(G_{t+1}^{(m)}-\mathbf1)\right).
\label{eq:myopic_eu}
\end{equation}
Here $c_t(w,\widetilde w_t)$ is the additive cost deduction in the one-period benchmark objective. Its realised performance is scored by the common multiplicative accounting in Equation~\eqref{eq:cost_adjusted_wealth}. This tests recurrent control against a one-period optimiser using the same dependence model.

The 2,048 benchmark scenarios differ from the 512 used for the RL risk charge. Solver convergence is checked. Appendix~A gives the numerical tolerances.

For the policy-information comparison, the common architecture activates the scalar scenario-CVaR observation, direct vine features, or both. Table~\ref{tab:information_comparison} compares these inputs, and Table~\ref{tab:curriculum_comparison} compares historical, synthetic, and mixed training.


\section{Empirical study}
\label{sec:empirical_study}

We evaluate NN-vine-RL, which combines neural-vine simulation with recurrent reinforcement learning. We first compare its portfolio performance with six benchmarks, then investigate the effects of training schemes, policy inputs, and portfolio aggregation.

\subsection{Data specifications}
\label{sec:data}
\label{subsec:4_description_data_metrics}

The universe contains three US price indices, three Chinese equity exchange-traded funds (ETFs), and a Chinese gold ETF. We obtain closing levels from Sina Finance through AKShare, with the instruments and source identifiers listed in Table~\ref{tab:assets}. The common calendar contains 2,946 closing levels and 2,945 daily log returns from 16 December 2013 to 6 July 2026.

\begin{table}[htb]
\centering
\caption{Seven-Asset Research Universe}
\label{tab:assets}
\small
\begin{tabular}{llll}
\toprule
Code & Identifier & Instrument & Quote \\
\midrule
SP500 & \texttt{.INX} & S\&P 500 price index & Index points \\
NASDAQ & \texttt{.IXIC} & NASDAQ Composite price index & Index points \\
DOW & \texttt{.DJI} & Dow Jones Industrial Average & Index points \\
SSE50 & \texttt{sh510050} & SSE 50 ETF & CNY \\
DIVIDEND & \texttt{sh510880} & Chinese dividend ETF & CNY \\
CHINEXT & \texttt{sz159915} & ChiNext ETF & CNY \\
GOLD & \texttt{sz159934} & Gold ETF & CNY \\
\bottomrule
\end{tabular}
\end{table}

The series are local-currency price returns. Distributions, currency conversion and FX hedging are not modelled. The common calendar aligns observed closing levels across exchange holidays, while US and Chinese closing times remain asynchronous. We compute daily log returns $r_{i,d}=\log P_{i,d}-\log P_{i,d-1}$ from these observations without interpolation or forward-filling. For a decision at $d_t$, holding-period gross returns compound the subsequent observations in $(d_t,d_{t+1}]$.

Table~\ref{tab:summary} shows negative skewness and substantial excess kurtosis for every asset. ChiNext is the most volatile series and gold the least volatile. The unconditional correlations in Table~\ref{tab:correlation} reveal strong US and Chinese equity blocks, weaker cross-market links and near-zero equity--gold correlation. Heavy tails motivate non-Gaussian marginal filters, while the block structure and the tail-sensitive control objective motivate modelling conditional dependence rather than relying on a constant covariance matrix.

\begin{table}[htb]
\centering
\caption{Daily Log-Return Descriptive Statistics, December 2013--July 2026}
\label{tab:summary}
\small
\resizebox{0.96\textwidth}{!}{%
\begin{tabular}{lrrrrrr}
\toprule
Asset & Mean (\%) & SD (\%) & Skewness & Excess kurtosis & Minimum (\%) & Maximum (\%) \\
\midrule
S\&P 500 & 0.0489 & 1.1112 & -0.631 & 16.145 & -12.765 & 9.089 \\
NASDAQ & 0.0635 & 1.3433 & -0.406 & 8.848 & -13.141 & 11.478 \\
Dow Jones & 0.0410 & 1.0892 & -0.819 & 22.081 & -13.842 & 10.764 \\
SSE 50 & 0.0218 & 1.3923 & -0.217 & 8.806 & -10.517 & 11.150 \\
Dividend ETF & 0.0187 & 1.3342 & -0.944 & 13.118 & -10.540 & 9.538 \\
ChiNext & 0.0391 & 2.1472 & -0.062 & 8.901 & -17.770 & 18.232 \\
Gold & 0.0448 & 0.9826 & -0.774 & 14.151 & -10.533 & 7.212 \\
\bottomrule
\end{tabular}}
\end{table}

\begin{table}[htb]
\centering
\caption{Unconditional Pearson Correlation Matrix}
\label{tab:correlation}
\small
\begin{tabular}{lrrrrrrr}
\toprule
 & SP500 & NASDAQ & DOW & SSE50 & DIV & CNXT & GOLD \\
\midrule
SP500 & 1.000 & 0.912 & 0.943 & 0.321 & 0.287 & 0.214 & -0.089 \\
NASDAQ & 0.912 & 1.000 & 0.876 & 0.298 & 0.262 & 0.198 & -0.072 \\
DOW & 0.943 & 0.876 & 1.000 & 0.314 & 0.279 & 0.207 & -0.093 \\
SSE50 & 0.321 & 0.298 & 0.314 & 1.000 & 0.857 & 0.694 & 0.052 \\
DIV & 0.287 & 0.262 & 0.279 & 0.857 & 1.000 & 0.621 & 0.048 \\
CNXT & 0.214 & 0.198 & 0.207 & 0.694 & 0.621 & 1.000 & 0.015 \\
GOLD & -0.089 & -0.072 & -0.093 & 0.052 & 0.048 & 0.015 & 1.000 \\
\bottomrule
\end{tabular}
\end{table}

The final 24 decisions, from 31 July 2024 to 30 June 2026, are excluded from model fitting and training. Requiring at least 15 common trading observations per holding period retains 22 periods and removes the intervals ending 27 February 2026 (14 common observations) and 6 July 2026 (three observations). The 22-return wealth curves are re-compounded after this selection.

Annualised volatility and Sharpe use the effective observations-per-year factor implied by the elapsed sample, with a zero annual risk-free rate. For simple net returns $R_t$ and $\gamma_U=2$, the reported annual certainty equivalent is
\begin{equation}
 \operatorname{CE}_{\rm ann}
 =\left[U_{\gamma_U}^{-1}\!\left\{n^{-1}\sum_{t=1}^n
 U_{\gamma_U}(1+R_t)\right\}\right]^{n/Y}-1,
 \label{eq:reported_annual_ce}
\end{equation}
where $Y$ is elapsed calendar years.

\subsection{Portfolio performance relative to benchmarks}
\label{subsec:4_outofsample}

Table~\ref{tab:outofsample} compares NN-vine-RL, an ensemble of ten independently trained policies, with six benchmarks on the common 22-period sample. All seven strategies use the same realised returns, feasible set, and cost model. The NN-vine benchmark uses the fitted neural-vine law for one-period portfolio optimisation.

NN-vine-RL records the highest observed total return (71.57\%), CAGR (34.13\%), Sharpe ratio (2.309), and annual CRRA certainty equivalent (33.07\%) in this comparison. Its 13.22\% annual volatility is lower than that of every actively optimised benchmark and close to equal weight. Relative to the static-vine benchmark, it adds 7.06 percentage points of total return and 3.52 points of annual CE while reducing monthly target-to-target turnover from 1.051 to 0.351.

\begin{table}[htb]
\centering
\caption{NN-vine-RL and Benchmarks on 22 Complete Periods.}
\label{tab:outofsample}
\resizebox{\textwidth}{!}{%
\begin{tabular}{l r r r r r r r}
\toprule
Strategy & Total return & CAGR & Volatility & Sharpe & Max DD & CRRA CE & Turnover \\
\midrule
Equal weight & 49.16\% & 24.37\% & 13.42\% & 1.702 & \textbf{5.22\%} & 23.38\% & \textbf{0.000} \\
Shrinkage mean--variance & 48.69\% & 24.16\% & 14.17\% & 1.608 & 7.83\% & 22.98\% & 0.052 \\
DCC--GARCH & 58.79\% & 28.69\% & 13.55\% & 1.944 & 5.79\% & 27.62\% & 0.153 \\
Static vine & 64.50\% & 31.19\% & 16.88\% & 1.704 & 6.51\% & 29.55\% & 1.051 \\
Rolling vine & 40.81\% & 20.52\% & 18.94\% & 1.082 & 9.90\% & 18.53\% & 1.237 \\
NN-vine & 61.44\% & 29.86\% & 16.45\% & 1.682 & 6.78\% & 28.30\% & 1.087 \\
\textbf{NN-vine-RL} & \textbf{71.57\%} & \textbf{34.13\%} & \textbf{13.22\%} & \textbf{2.309} & 5.84\% & \textbf{33.07\%} & 0.351 \\
\bottomrule
\end{tabular}}
\end{table}

Equal weight's zero entry is zero \emph{target-to-target} turnover: constant target weights are treated as unchanged at monthly decisions.

\begin{figure}[htb]
\centering
\includegraphics[width=481.52988pt,height=323.98558pt]{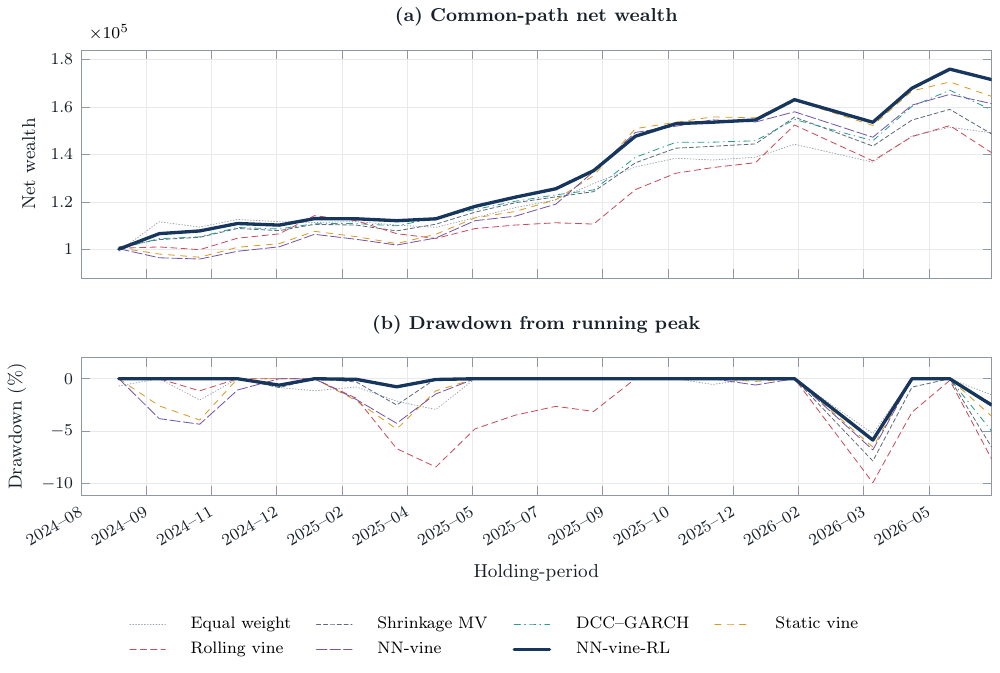}
\caption{Net wealth and drawdown re-compounded from the 22 retained net returns, with initial wealth of 100,000. The two incomplete holding periods do not enter either path.}
\label{fig:oos-wealth-drawdown}
\end{figure}

\begin{figure}[htb]
\centering
\includegraphics[width=431.66681pt,height=200.668pt]{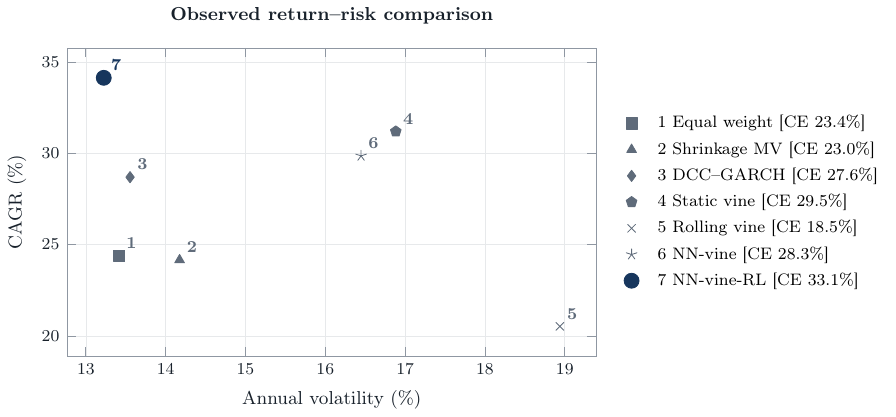}
\caption{Observed CAGR--volatility positions and annual CRRA certainty equivalents on the common 22-period sample.}
\label{fig:primary_risk_return}
\end{figure}

Figures~\ref{fig:oos-wealth-drawdown} and \ref{fig:primary_risk_return} show that NN-vine-RL achieves higher growth and CRRA certainty equivalent than equal weight, with similar volatility. Equal weight has the smallest drawdown. The myopic NN-vine optimiser also performs strongly, while NN-vine-RL has substantially lower target-to-target turnover under the common accounting convention. We next examine which training and information choices generate that behaviour.

\subsection{Ablation study}
\label{sec:causal_component_analysis}

We examine how training scenarios and policy inputs contribute to the performance of the NN-vine-RL ensemble in Section~\ref{subsec:4_outofsample}. The reference model combines mixed pre-training and historical fine-tuning with a compressed policy state. We first vary the training scheme while fixing this state, then vary the policy inputs while fixing the mixed training scheme. Both comparisons use ten matched optimiser seeds, the common 22-period evaluation sample, and the same portfolio constraints and cost accounting. We average the ten policies' target weights before computing ensemble returns and costs.

\paragraph{Training schemes.}
We compare mixed pre-training followed by historical fine-tuning, historical training throughout, synthetic pre-training followed by historical fine-tuning, and synthetic training throughout. All four schemes use the compressed state. \hyperref[subsec:7_numerical_implementation_reproducibility]{Appendix A} gives the mixed scheme's training budget.

Table~\ref{tab:curriculum_comparison} shows that mixing vine-simulated paths with historical training maintains strong return performance while substantially reducing trading. Relative to historical-only training, the mixed scheme raises CAGR from 33.05\% to 34.13\% and annual CE from 32.18\% to 33.07\%, while reducing monthly turnover from 0.600 to 0.351, a 41.5\% decrease. Historical-only training achieves slightly lower volatility and drawdown and a higher Sharpe ratio. The two schemes also deliver closely matched CE over all 24 holding periods, at 30.38\% for mixed training and 30.51\% for historical-only training. On the common 22-period sample, synthetic pre-training followed by historical fine-tuning retains a similar low-turnover profile, with CE of 30.90\%, while synthetic-only training yields CE of 23.55\%.

\begin{table}[htb]
\centering
\caption{Training Schemes for NN-vine-RL}
\label{tab:curriculum_comparison}
\small
\begin{tabular}{lrrrrrr}
\toprule
Training scheme & CAGR & Volatility & Sharpe & Max DD & CRRA CE & Turnover \\
\midrule
\textbf{Mixed + FT} & \textbf{34.13\%} & 13.22\% & 2.309 & 5.84\% & \textbf{33.07\%} & \textbf{0.351} \\
Historical only & 33.05\% & \textbf{12.12\%} & \textbf{2.441} & \textbf{4.76\%} & 32.18\% & 0.600 \\
Synthetic + FT & 32.26\% & 15.14\% & 1.939 & 7.46\% & 30.90\% & 0.354 \\
Synthetic only & 24.63\% & 13.78\% & 1.677 & 7.23\% & 23.55\% & 0.365 \\
\bottomrule
\end{tabular}
\end{table}

We use paired circular-block resampling for the three mixed-minus-comparator CE contrasts \autocite{politisRomano1992circular}, with 9,999 replications of block length three. Table~\ref{tab:mixed_inference} reports pointwise intervals and one-sided tests with Holm adjustment \autocite{holm1979multiple}.

The Holm-adjusted $p$-values range from 0.1077 to 0.6406. Mixed training improves seed-level CE in eight of ten comparisons with synthetic-only training, with a mean gain of 8.85 percentage points. Against historical-only training, the mean seed-level CE difference is 0.24 percentage points. Together, these comparisons identify the main benefit of mixing vine-simulated and historical paths as a substantial reduction in turnover with maintained growth and utility performance.

\begin{table}[htb]
\centering
\caption{CE Differences (Mixed Minus Comparator) and Matched-Seed Effects. *Intervals are pointwise}
\label{tab:mixed_inference}
\small
\begin{tabular}{lrrrrrrr}
\toprule
Comparator & CE diff. & 95\% interval* & One-sided $p$ & Holm $p$ & Positive diff. & Mean & Median \\
\midrule
Historical only & 0.89 pp & $[-8.49,\ 14.88]$ pp & 0.4172 & 0.6406 & 4/10 & 0.24 pp & $-2.81$ pp \\
Synthetic only & 9.52 pp & $[0.58,\ 20.09]$ pp & 0.0359 & 0.1077 & 8/10 & 8.85 pp & 8.53 pp \\ 
Synthetic + FT & 2.17 pp & $[-7.57,\ 10.90]$ pp & 0.3203 & 0.6406 & 5/10 & 1.70 pp & 0.25 pp \\
\bottomrule
\end{tabular}
\end{table}

\begin{figure}[htb]
\centering
\includegraphics[width=462.58124pt,height=175.51797pt]{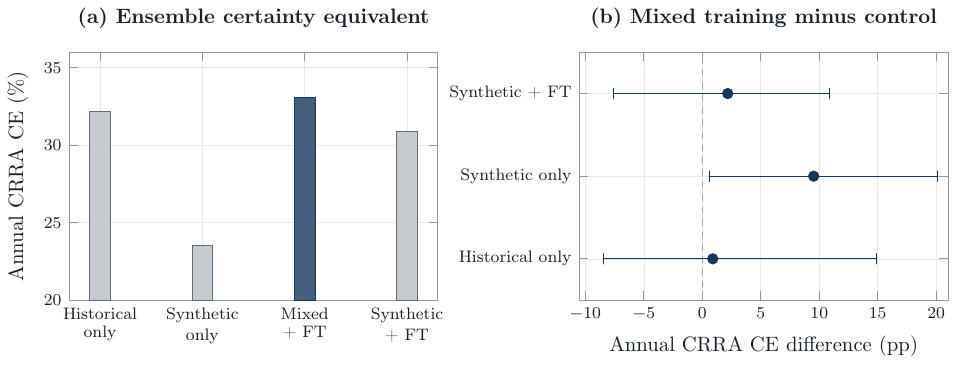}
\caption{Training schemes for NN-vine-RL under the compressed state. FT denotes historical fine-tuning after pre-training. Historical only and Synthetic only use a single trajectory source throughout training. Panel (a) reports ensemble annual CRRA CE in percent. Panel (b) reports mixed-minus-control differences in percentage points with pointwise 95\% circular-block intervals. Family-adjusted tests are given in Table~\ref{tab:mixed_inference}.}
\label{fig:mixed_pretraining_comparison}
\end{figure}
\vspace{-15pt}

\begin{samepage}
\paragraph{Policy inputs.}
We next hold mixed pre-training and historical fine-tuning fixed and vary the information supplied directly to the policy. The compressed state sets the scalar scenario-CVaR observation and the 63 direct vine coordinates to zero. We compare it with three variants that expose the scalar observation, the vine coordinates, or both. All variants retain the same 88-dimensional input schema, recurrent architecture, training paths, scenario-risk charge, and ten optimiser seeds. The compressed-state row in Table~\ref{tab:information_comparison} uses the same ten policies as the main model in Table~\ref{tab:outofsample}.
\par
\end{samepage}

\begin{table}[htb]
\centering
\caption{Policy Inputs for NN-vine-RL}
\label{tab:information_comparison}
\small
\begin{tabular}{lrrrrrr}
\toprule
Policy inputs & CAGR & Volatility & Sharpe & Max DD & CRRA CE & Turnover \\
\midrule
\textbf{Compressed state} & \textbf{34.13\%} & 13.22\% & \textbf{2.309} & \textbf{5.84\%} & \textbf{33.07\%} & \textbf{0.351} \\
Scalar scenario-CVaR & 26.01\% & \textbf{12.76\%} & 1.890 & 6.33\% & 25.05\% & 0.408 \\
Raw vine coordinates & 32.45\% & 14.24\% & 2.064 & 6.42\% & 31.23\% & 0.458 \\
Full state & 30.78\% & 13.26\% & 2.109 & 6.36\% & 29.73\% & 0.406 \\
\bottomrule
\end{tabular}
\end{table}

The compressed state achieves the highest CE and lowest turnover among the four input specifications in Table~\ref{tab:information_comparison}. Its CE exceeds those of the scalar-CVaR, raw-vine, and full-state variants by 8.02, 1.84, and 3.34 percentage points, respectively.

Using the same circular-block resampling settings as for the training comparison, the pointwise 95\% intervals for these three CE differences are $[-1.30,18.62]$, $[-2.73,6.66]$, and $[-2.53,10.68]$ percentage points. Holm-adjusted two-sided $p$-values are 0.3198, 0.6050, and 0.6050, respectively. The compressed state has higher seed-level CE in seven of ten comparisons with the scalar-CVaR variant and five of ten comparisons with each of the other variants. These results favour the compressed state for its combination of return, risk, and trading intensity. The neural vine contributes to this policy through simulated training paths and the scenario-risk charge, while the policy conditions directly on wealth, returns, volatilities, and portfolio holdings.

\paragraph{Portfolio aggregation.}

NN-vine-RL averages the target weights of its ten independently trained policies before applying realised returns and costs. Across the 22 evaluation periods, average gross exposure falls from 1.399 for the individual policies to 1.059 for the ensemble. Averaging cancels 85.31\% of the incremental gross exposure above one, with the same proportional reduction in short exposure. Costs reduce the ensemble's total return by 1.61 percentage points.

Weight averaging offsets opposing long and short positions before transaction and financing costs are charged. This aggregation complements the mixed training scheme and compressed policy state by keeping the implemented portfolio's gross exposure close to one.

\section{Conclusion}
\label{sec:conclusion}

In this paper, we study dynamic portfolio selection under a CRRA--CVaR objective using conditional vine simulation and recurrent reinforcement learning. The fitted return law supplies both training paths and scenarios for evaluating portfolio risk. Under uniform local derivative bounds, we propagate pair-copula parameter errors through the conditional transforms and obtain Hellinger and total variation bounds of order $\varepsilon^{1/(2d-1)}$. Together with terminal-utility moment assumptions, these estimates yield bounds on policy values, optimal values and the true-market performance loss of policies selected under the fitted model. The resulting performance bound separates marginal and structural misspecification, pair-copula parameter error, simulation error and numerical optimisation. Our numerical study illustrates the practical effectiveness of the approach using a neural Student-$t$ D-vine and recurrent TD3.

Several directions remain for future research. These include deriving sharper family-specific stability bounds, analysing degenerate parameter limits and verifying the local derivative condition for further copula families. Quantifying the simulation and optimisation errors of the policy-learning algorithm and adapting the vine and recurrent policy architectures to real-time market updates are further directions.

\pdfbookmark[1]{Appendices}{appendices}
\section*{Appendices}
\label{sec:appendix}

Appendix~A records the numerical configuration. Appendix~B reports generative-fidelity and training diagnostics. Appendix~C provides proofs and supplementary derivations for Sections~\ref{sec:problem_formulation} and \ref{sec:dependence_robustness}.

\phantomsection
\subsection*{A. Numerical implementation and configuration}
\label{subsec:7_numerical_implementation_reproducibility}

\setcounter{figure}{0}
\renewcommand{\thefigure}{A.\arabic{figure}}
\setcounter{table}{0}
\renewcommand{\thetable}{A.\arabic{table}}

Table \ref{tab:hyperparameters} summarizes the detailed numerical configuration used throughout the empirical study.

We use R for marginal and vine estimation, accounting, and benchmark optimisation, and PyTorch for policy learning and inference. Each vine-edge network has two hidden layers with widths 16 and 8.

The main model comprises ten independently trained policies with the compressed state described in Section~\ref{sec:causal_component_analysis}. Each policy receives 1,000 mixed pre-training presentations: 621 draws from 100 distinct neural-vine paths and 379 draws from 61 distinct historical paths. Historical fine-tuning then makes one pass through all 61 historical paths. Appendix~B reports diagnostics for these synthetic paths and the mixed pre-training stage of these ten policies.

\begin{table}[htbp]
\centering
\caption{Numerical configuration}
\label{tab:hyperparameters}
\small
\resizebox{0.98\textwidth}{!}{%
\begin{tabular}{lll@{\qquad}lll}
\toprule
\textbf{Quantity} & \textbf{Symbol} & \textbf{Value} &
\textbf{Quantity} & \textbf{Symbol} & \textbf{Value} \\
\midrule
Sequence length & $L$ & 30 & Observation dimension & $d_o$ & 88 \\
LSTM units/layers & --- & 64/1 & Replay capacity & $N_B$ & 100,000 \\
Actor LR (pre/fine) & $\alpha_\mu$ & $3\!\times\!10^{-5}/2\!\times\!10^{-5}$ & Critic LR & $\alpha_Q$ & $10^{-4}$ \\
Batch (pre/fine) & $B$ & 128/64 & Random warm-up & --- & 1,000 steps \\
Discount/target rate & $\gamma,\tau$ & $1,0.005$ & Policy delay & $d$ & 2 \\
Target noise/clip & $\sigma,c$ & $0.10,0.25$ & Gradient-norm cap & --- & 1 \\
CRRA coefficient & $\gamma_U$ & 2 & CVaR level/penalty & $\alpha,\lambda$ & $0.95,0.05$ \\
Scenarios per step & $M$ & 512 & Turnover cost & $\kappa$ & 0.001 \\
Gross/net limits & --- & $1.5/1.0$ & Position bounds & --- & $[-0.20,0.60]$ \\
\bottomrule
\end{tabular}}
\end{table}

We use Adam \autocite{kingmaAdamMethodStochastic2015} for policy learning. Benchmark optimisation uses tolerance $10^{-9}$ and at most 2,000 evaluations.

\phantomsection
\subsection*{B. Generative fidelity and training dynamics}
\label{subsec:7_generative_fidelity_training_dynamics}

\setcounter{figure}{0}
\renewcommand{\thefigure}{B.\arabic{figure}}
\setcounter{table}{0}
\renewcommand{\thetable}{B.\arabic{table}}

\begin{figure}[p]
\centering
\includegraphics[width=468.0087pt,height=595.4171pt]{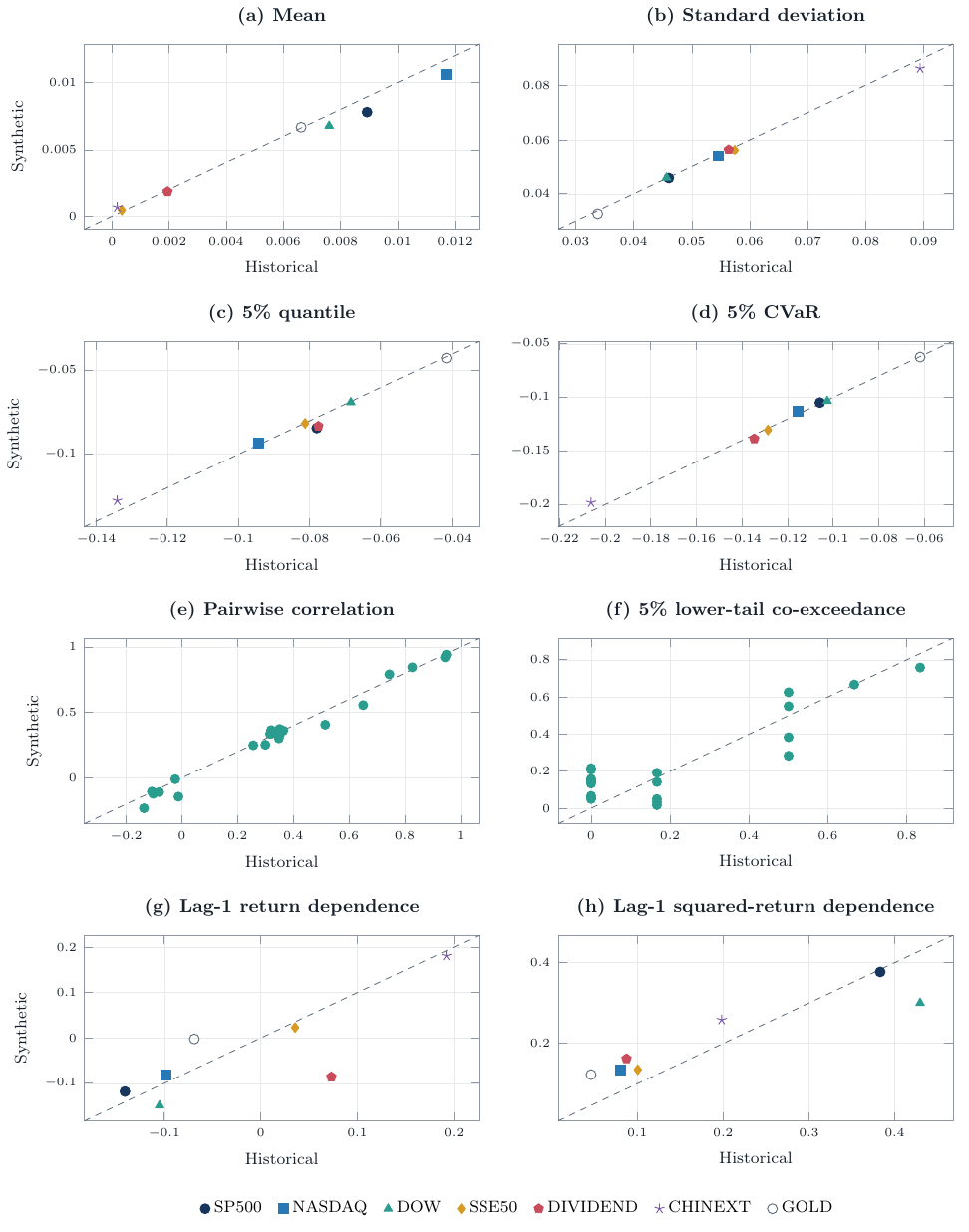}
\caption{Generative fidelity of the 100 synthetic paths used by the main model. Panels (a)--(d) compare marginal moments and tail statistics of monthly log returns. Panel (e) reports pairwise correlations, and panel (f) reports conditional lower-tail co-exceedance using each asset's 5\% quantile. Panels (g)--(h) compare first-order temporal dependence, using adjacent observations within each synthetic path. The grey diagonal denotes exact agreement.}
\label{fig:appendix_synthetic_fidelity_master}
\end{figure}

We examine the scenario sample and the ten compressed-state policies used by the main model in Section~\ref{subsec:4_outofsample}. Figure~\ref{fig:appendix_synthetic_fidelity_master} compares the 2,400 realised monthly log returns in the 100 synthetic paths with the 114 distinct historical monthly observations in the training prefix. Panels (a)--(d) report marginal means, standard deviations, 5\% quantiles, and lower-tail means at the 5\% level (CVaR). Panels (e)--(h) compare pairwise correlations, lower-tail co-exceedance, and lag-one dependence in returns and squared returns. Synthetic lag statistics use the 2,300 adjacent-month pairs within the 24-month paths.

The synthetic paths closely reproduce the marginal distributions. Mean differences are at most 0.025 historical standard deviations, and standard-deviation errors range from $-3.64\%$ to $0.56\%$. The largest differences in 5\% quantiles and lower-tail means are 0.138 and 0.090 historical standard deviations, respectively. Across the 21 asset pairs, the mean absolute correlation difference is 0.039, with 19 differences below 0.10. Temporal dependence varies across assets, with maximum absolute differences of 0.159 for returns and 0.129 for squared returns.

\begin{figure}[htbp]
\centering
\includegraphics[width=451.16281pt,height=277.96983pt]{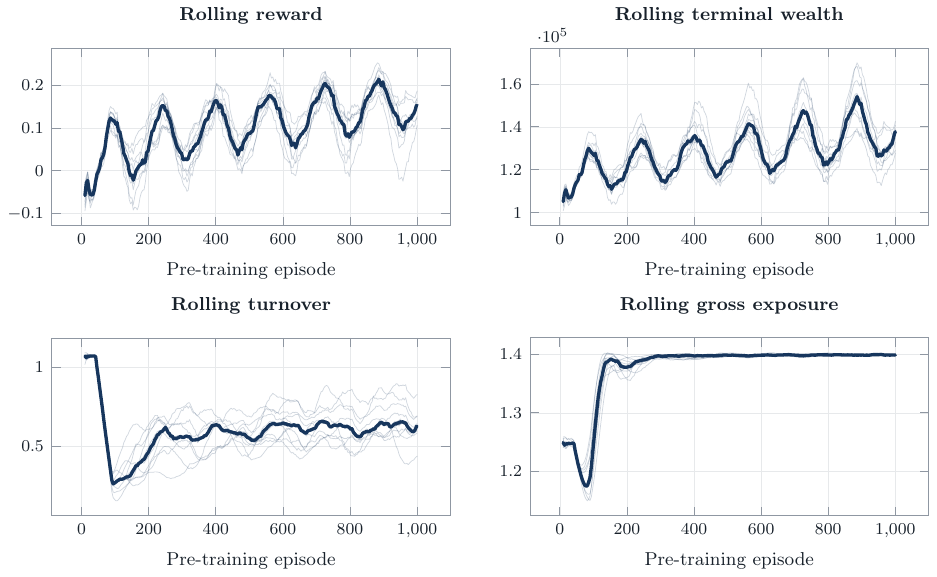}
\caption{Mixed pre-training behaviour of the ten compressed-state policies used by the main model. Each curve is a 50-episode trailing mean, shown from the tenth episode using the available observations. Thin transparent lines identify individual seeds, and the dark line is the cross-seed median.}
\label{fig:appendix_pretraining_stability}
\end{figure}

\begin{figure}[htbp]
\centering
\includegraphics[width=445.54413pt,height=274.66312pt]{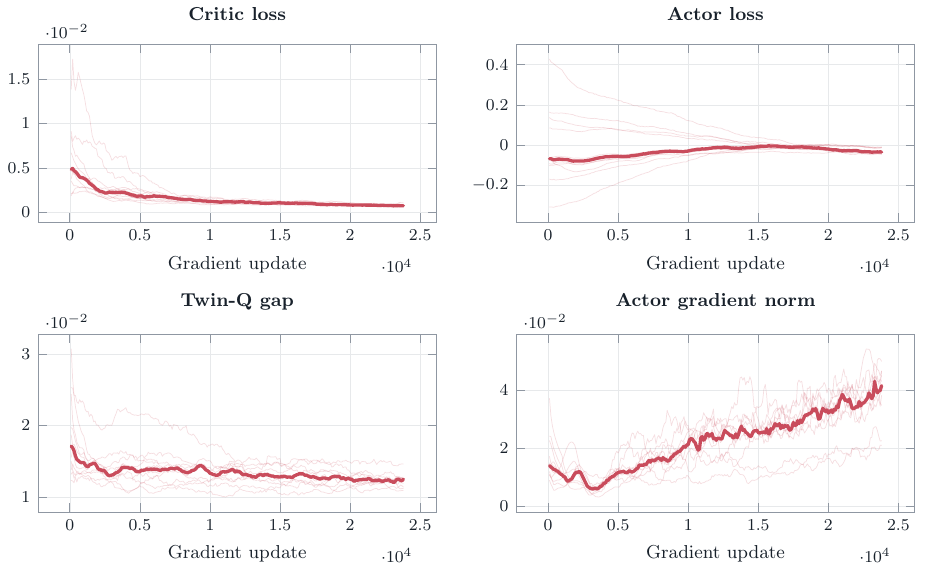}
\caption{Mixed pre-training optimiser diagnostics for the ten compressed-state policies used by the main model. Curves are trailing means over ten recorded observations, using the available observations at the start, and are plotted against gradient updates. Diagnostics are recorded every 100 updates. Thin transparent lines identify individual seeds, and the dark line is the cross-seed median.}
\label{fig:appendix_optimizer}
\end{figure}

Figures~\ref{fig:appendix_pretraining_stability} and \ref{fig:appendix_optimizer} report the mixed pre-training stage with the fitted vine model held fixed. Episode reward and terminal wealth fluctuate over the mixed presentations and increase overall. Comparing the first and last 50 episodes, the cross-seed median of mean terminal wealth rises from 114,234 to 137,740, while mean turnover falls from 0.933 to 0.627. Gross exposure stabilises near 1.40. Critic loss declines during training, the dispersion of actor loss across seeds narrows, and the twin-critic gap stabilises in the later updates. The actor gradient norm continues to increase gradually towards the end of pre-training.

\pdfbookmark[2]{C. Proofs}{appendix.C}
\phantomsection
\subsection*{C. Proofs}
\label{app:dependence_proofs}

We present proofs of results in Sections~\ref{sec:problem_formulation} and \ref{sec:dependence_robustness} and also verify the regular copula examples. Recall the common state and wealth dynamics in Equation~\eqref{eq:ideal_market_dynamics}.

\setcounter{theorem}{0}
\renewcommand{\thetheorem}{C.\arabic{theorem}}
\renewcommand{\theHtheorem}{C.\arabic{theorem}}
\setcounter{equation}{0}
\renewcommand{\theequation}{C.\arabic{equation}}
\renewcommand{\theHequation}{C.\arabic{equation}}

\setcounter{subsubsection}{0}
\renewcommand{\thesubsubsection}{C.\arabic{subsubsection}}
\renewcommand{\theHsubsubsection}{C.\arabic{subsubsection}}

\subsubsection{Proofs for Section~\ref{sec:problem_formulation}}
\label{app:problem_formulation_proofs}
\phantomsection\label{app:reachable_regularity_proof}
\begin{proof}[Proof of Lemma~\ref{lem:reachable_regularity}]
Iterating the wealth recursion from $X_0=1$ and using Equation~\eqref{eq:ideal_solvency} proves the positive wealth lower bound in Equation~\eqref{eq:market_state_bounds}.

The initial state space $\mathcal S_0$ is a singleton. Continuity and compact values give each $A_j$ a closed graph, and all actions lie in the compact ball $\{a:\|a\|_1\leq\Lambda\}$. Continuity of $c_j$ and the positive lower bound on $K_j$ make $\Phi_j$ continuous on its feasible domain. To prove relative closedness, let $s_t^n\in\mathcal S_t$ converge to $s_t$ in $(\{H_0\}\times\mathcal G^t)\times(0,\infty)\times\mathbb R^d$. The history component gives convergence of each return in the generating paths. Choose a feasible generating action sequence $(a_0^n,\ldots,a_{t-1}^n)$ for each $s_t^n$. The uniform action bound gives a subsequence along which all these actions converge. Starting from $s_0$, the closed graph of $A_0$ makes the limiting first action feasible, and continuity of $\Phi_0$ gives convergence of the first states to a state in $\mathcal S_1$. Repeating this argument at each date establishes membership of each limiting state in $\mathcal S_j$ before applying the closed graph of $A_j$. Every limiting action is therefore feasible, and the resulting terminal state is $s_t$. Thus $s_t\in\mathcal S_t$ by Equation~\eqref{eq:reachable_state_spaces}. Each $\mathcal S_t$ is relatively closed in a Polish space and hence is Polish. Nonemptiness follows from nonemptiness of $A_t(s)$ and $\mathcal G$. Induction completes the proof.
\end{proof}

\phantomsection\label{app:dp_proof}
\begin{proof}[Proof of Proposition~\ref{prop:dynamic_programming}]
By Lemma~\ref{lem:reachable_regularity}, the reachable state spaces are Polish, $\Phi_t$ is continuous, and $X\geq\underline k^{\,t}>0$. Hence $u_t$ and $\ell_t=1-K_t$ are continuous on the feasible state--action--return domain.

Set $\ell_+=1-\underline k$ and define the common threshold interval
\begin{equation}
 I_v=\left[1-\underline k-\frac{B-\underline k}{\alpha},\;1-\underline k\right].
 \label{eq:cvar_threshold_interval}
\end{equation}
For every state, action, and model, the infimum in the CVaR representation can be restricted to $I_v$. Indeed, if $B>\underline k$, the nonnegative variable $Y=K_t-\underline k$ satisfies $\mathbb E[Y\mid H]\leq B-\underline k$, so Markov's inequality gives
\begin{equation*}
 \mathbb P\!\left(\ell_t<\ell_+-\frac{B-\underline k}{\alpha}\,\middle|\,H\right)
 =\mathbb P\!\left(Y>\frac{B-\underline k}{\alpha}\,\middle|\,H\right)
 \leq\alpha.
\end{equation*}
Since $\ell_t\leq\ell_+$, an $\alpha$-quantile, and hence a minimising threshold, lies in $I_v$. If $B=\underline k$, then $K_t=\underline k$ almost surely under each return kernel and $I_v=\{\ell_+\}$ contains a minimiser. On this compact threshold interval,
\begin{equation*}
 0\leq(\ell_t-v)_+\leq\frac{B-\underline k}{\alpha},\qquad v\in I_v.
\end{equation*}
The integrand is therefore bounded and jointly continuous in the state, action, return, and threshold. Weak continuity of the return kernel makes
\begin{align*}
(s,a,v)\longmapsto
v+\frac{1}{p}\mathbb E[(\ell_t(s,a,G)-v)_+\mid s]
\end{align*}
continuous. Taking its minimum over the compact threshold interval proves joint continuity of $\mathcal C_t(s,a)$ in $(s,a)$. Moreover,
\begin{equation*}
 1-B\leq\mathbb E[\ell_t\mid H]\leq\mathcal C_t(s,a)\leq1-\underline k.
\end{equation*}

Starting from $V_T=0$, suppose that $V_{t+1}$ is continuous. We verify the integrability and growth bounds needed for the induction. If $\gamma_U>1$, Equation~\eqref{eq:market_state_bounds} gives
\begin{equation*}
 |u_t(s,a,g)|\leq
 \frac{\underline k^{\,t(1-\gamma_U)}+\underline k^{\,(t+1)(1-\gamma_U)}}{\gamma_U-1}.
\end{equation*}
Thus backward induction gives bounded value functions. If $0<\gamma_U<1$, set $r=1-\gamma_U\in(0,1)$ and use the induction bound $|V_{t+1}(s)|\leq C_{t+1}(1+X^r)$, with a finite constant $C_{t+1}$. Jensen's inequality and the first-moment bound give
\begin{align*}
 \mathbb E[K_t^r\mid H]&\leq B^r,\\
 \mathbb E\!\left[K_t^r\mathbf1_{\{K_t>R\}}\,\middle|\,H\right]
 &\leq BR^{r-1},\qquad R>0.
\end{align*}
Since $|u_t|\leq X^r(K_t^r+1)/r$ and the next wealth is $XK_t$, we obtain
\begin{align*}
 |Q_t(s,a)|
 &\leq \frac{X^r(B^r+1)}{r}
       +C_{t+1}(1+X^rB^r)+\lambda\max\{|1-B|,|1-\underline k|\}\\
 &\leq C_t(1+X^r),
 \qquad |V_t(s)|\leq C_t(1+X^r),
\end{align*}
for a finite constant $C_t$. Along any convergent feasible state--action sequence, $X$ is bounded, so the tail estimate gives uniform integrability of the Bellman integrands. If $\gamma_U=1$, then $u_t=\log K_t$, its negative part is bounded by $\max\{0,-\log\underline k\}$, and its positive part satisfies
\begin{equation*}
 \mathbb E\!\left[\log^+K_t\,\mathbf1_{\{K_t>R\}}\,\middle|\,H\right]
 \leq B\sup_{x>R}\frac{\log x}{x}\longrightarrow0
 \qquad(R\to\infty,\ R\geq1).
\end{equation*}
Consequently, backward induction gives bounded value functions and uniformly integrable Bellman integrands. Continuity of $\Phi_t$ and weak continuity of the return kernel, applied first to bounded continuous truncations and then using these tail estimates, imply continuity of
\begin{align*}
(s,a)\longmapsto
\mathbb E\!\left[
u_t(s,a,G)+V_{t+1}(\Phi_t(s,a,G))\mid s\right].
\end{align*}
Thus $Q_t$ is finite and continuous. The maximum theorem applied to $Q_t$ and $A_t$ gives a finite continuous $V_t$ and nonempty compact sets
\begin{equation*}
 A_t^*(s)=\{a\in A_t(s):Q_t(s,a)=V_t(s)\}.
\end{equation*}
The graph of $A_t^*$ is Borel and its sections are nonempty and compact, so it admits a Borel selector $\pi_t^*$. For any admissible $\pi$, the growth bounds and the tower property give Equation~\eqref{eq:fixed_policy_bellman}. Backward induction yields finite Borel $V_t^\pi$, continuous when $\pi$ is continuous. The same induction applies to the fitted model with the same constants.

Finally, for any admissible policy $\pi$, the conditional return law and the Bellman maximum give
\begin{align*}
 &\mathbb E^\pi\!\left[
 r_t(s_t,\pi_t(s_t),G_{t+1})+V_{t+1}(s_{t+1})
 \,\middle|\,\mathcal F_t\right]\\
 &\qquad=Q_t(s_t,\pi_t(s_t))\leq V_t(s_t).
\end{align*}
Taking expectations and summing over $t$, with $V_T=0$, yields
\begin{equation*}
 J(\pi;s_0)\leq V_0(s_0)=J(\pi^*;s_0),
\end{equation*}
where equality holds because $\pi_t^*$ maximises $Q_t$ at every state. The Borel selectors and state recursion make $\pi_t^*(s_t)$ adapted to $\mathcal F_t$, so $\pi^*$ is admissible and optimal.
\end{proof}

\subsubsection{Proofs for Section~\ref{sec:dependence_robustness}}
\label{app:dependence_results_proofs}
\phantomsection\label{app:vine_regularity_examples}
\begin{proof}[Proof of Example~\ref{ex:regular_pair_copulas}]
The standard densities and conditional transforms of these families are
strictly positive and smooth on interior inputs, with the conditional
normalisations in Assumption~\ref{ass:vine_regular}. We verify the derivative
bounds and the Clayton endpoint. Below, $C$ depends only on the stated
compact parameter domain and may change between estimates. Symmetry gives
the same bounds for both orientations.

\emph{Gaussian.}
Let $z=\Phi^{-1}(u)$, $w=\Phi^{-1}(v)$ and
$a=(z-\rho w)/\sqrt{1-\rho^2}$. Then
\begin{align*}
 \log c_\rho(u,v)
 &=-\frac12\log(1-\rho^2)
   +\frac{2\rho zw-\rho^2(z^2+w^2)}{2(1-\rho^2)},\\
 h_\rho(u\mid v)&=\Phi(a).
\end{align*}
On $[\tau,1-\tau]^2$, $|z|,|w|\le a_\tau=\Phi^{-1}(1-\tau)$,
where $a_\tau^2\le C\log(e/\tau)$ and
$a_\tau/\phi(a_\tau)\le1/\tau$ by the normal tail bound.
Differentiating the two formulas, using
$\phi(z),\phi(w)\ge\phi(a_\tau)$ and $1-\rho^2\ge1-\rho_*^2$, gives
\begin{equation*}
 L_e(\tau)\le C\max\left\{1+a_\tau^2,
                    \frac{1+a_\tau}{\phi(a_\tau)}\right\}\le\frac C\tau.
\end{equation*}

\emph{Student-$t$.}
Write $x=T_\nu^{-1}(u)$, $y=T_\nu^{-1}(v)$ and
$Q_\rho=(x^2-2\rho xy+y^2)/(1-\rho^2)$.
The density ratio in \textcite{demartaTCopulaRelated2005} gives
$\log c_{\rho,\nu}(u,v)=\ell_{\rho,\nu}(x,y)$, where
\begin{align*}
 \ell_{\rho,\nu}(x,y)
 ={}&\log\frac{\Gamma((\nu+2)/2)\Gamma(\nu/2)}
                         {\Gamma((\nu+1)/2)^2}
 -\frac12\log(1-\rho^2)-\frac{\nu+2}{2}\log(1+Q_\rho/\nu)\\
 &+\frac{\nu+1}{2}\{\log(1+x^2/\nu)+\log(1+y^2/\nu)\}.
\end{align*}
Since $\nu+Q_\rho$ is uniformly comparable to $1+x^2+y^2$,
direct differentiation at fixed $x,y$ gives
\begin{align*}
 |\partial_x\ell|&\le\frac C{1+|x|},\qquad
 |\partial_y\ell|\le\frac C{1+|y|},\qquad
 |\partial_\rho\ell|\le C,\\
 |\partial_\nu\ell|&\le C\{\log(e+|x|)+\log(e+|y|)\}.
\end{align*}
The univariate density satisfies
$f_\nu(x)\asymp(1+|x|)^{-\nu-1}$ and
$|\partial_\nu\log f_\nu(x)|\le C\log(e+|x|)$, uniformly in $\nu$.
Integrating these bounds over the appropriate tail yields
\begin{align*}
 \frac{T_\nu(x)[1-T_\nu(x)]}{f_\nu(x)}&\le C(1+|x|),\\
 \frac{|\partial_\nu T_\nu(x)|}{f_\nu(x)}
 &\le C\frac{\int_{|x|}^\infty f_\nu(r)\log(e+r)\,dr}{f_\nu(x)}
 \le C(1+|x|)\log(e+|x|).
\end{align*}
The score bound supplies an integrable majorant for differentiating the
tail integral. Implicit differentiation of $T_\nu(x)=u$ therefore gives
$|\partial_\nu x|\le C(1+|x|)\log(e+|x|)$.
The same holds for $y$, and the polynomial tails imply
$\log(e+|x|)+\log(e+|y|)\le C\log(e/\tau)$ on $[\tau,1-\tau]^2$.
The chain rule now gives
\begin{align*}
 |\partial_\rho\log c|+|\partial_\nu\log c|&\le C\log(e/\tau),\\
 u(1-u)|\partial_u\log c|+v(1-v)|\partial_v\log c|&\le C.
\end{align*}
For the conditional transform, use
\begin{equation*}
 h_{\rho,\nu}(u\mid v)=T_{\nu+1}(z),\qquad
 z=\frac{x-\rho y}{s},\qquad
 s^2=\frac{(\nu+y^2)(1-\rho^2)}{\nu+1}.
\end{equation*}
Here $s\asymp1+|y|$ and $(1+|x|)/s\le C(1+|z|)$.
Differentiation gives
\begin{equation*}
 \partial_u z=\frac1{s f_\nu(x)},\qquad
 \partial_v z=\frac{-\rho/s-zy/(\nu+y^2)}{f_\nu(y)},\qquad
 \partial_\rho z=-\frac ys+\frac{\rho z}{1-\rho^2}.
\end{equation*}
Combining these formulas with the tail ratio above bounds
$u(1-u)|\partial_u z|+v(1-v)|\partial_v z|+|\partial_\rho z|$
by $C(1+|z|)$. The quantile derivative bounds similarly give
$|\partial_\nu z|\le C(1+|z|)\log(e/\tau)$.
Since
\begin{equation*}
 \sup_{\nu,z}\bigl\{(1+|z|)f_{\nu+1}(z)
                    +|\partial_\nu T_{\nu+1}(z)|\bigr\}\le C,
\end{equation*}
the chain rule for $T_{\nu+1}(z)$ gives the same weighted input bounds
and logarithmic parameter bounds for $h$.
Thus $L_e(\tau)\le C/\tau$.

\emph{Clayton.}
For $\theta>0$, put $x=-\log u$, $y=-\log v$ and
$G(\theta)=\log(e^{\theta x}+e^{\theta y}-1)$. Then
\begin{align*}
 \log c_\theta&=\log(1+\theta)+(1+\theta)(x+y)-2G(\theta)-G(\theta)/\theta,\\
 \log h_\theta(u\mid v)&=(1+\theta)y-G(\theta)-G(\theta)/\theta.
\end{align*}
We have $G(0)=0$, $|G'|\le x+y$ and $|G''|\le2(x+y)^2$.
The identities
\begin{equation*}
 \frac{G(\theta)}\theta=\int_0^1G'(r\theta)\,dr,\qquad
 \partial_\theta\frac{G(\theta)}\theta=\int_0^1rG''(r\theta)\,dr
\end{equation*}
give joint continuously differentiable extensions at $\theta=0$, with
$c_0=1$ and $h_0(u\mid v)=u$. They also yield
\begin{equation*}
 |\partial_\theta\log c_\theta|+|\partial_\theta h_\theta|
 \le C\{1+(x+y)^2\}\le C[\log(e/\tau)]^2.
\end{equation*}
For the input derivatives, $e^{G(\theta)}\ge u^{-\theta},v^{-\theta}$
gives
\begin{equation*}
 |\partial_u\log c_\theta|+|\partial_u h_\theta|\le C/u,\qquad
 |\partial_v\log c_\theta|+|\partial_v h_\theta|\le C/v.
\end{equation*}
Since $\sup_{0<\tau<1/4}\tau[\log(e/\tau)]^2<\infty$, these parameter
and input estimates give $L_e(\tau)\le C/\tau$.

Taking the maximum of the constants over the finite edge set and
increasing it to at least one gives the required $A$.
Measurability follows from the assumed measurable parameter functions.
\end{proof}

\phantomsection\label{app:vine_density_proof}
\begin{proof}[Proof of Theorem~\ref{thm:vine_density_stability}]

Fix a date $t$ and raw market history $H$, and suppress these indices.
We interpolate the pair parameters, localise the resulting square-root
densities, and control the probability removed by this localisation.

\emph{Parameter interpolation and conditional inputs.}
For every edge $e$, define
\begin{equation*}
 \theta_e^s=\theta_e^\circ+s(\widehat\theta_e-\theta_e^\circ),
 \qquad 0\le s\le1.
\end{equation*}
Convexity of $\Theta_e$ makes these valid pair parameters.
Let $c_s$ be the resulting full vine density and $U^s_{i\mid D}$ its
conditional uniforms, computed along the fixed recursion.
We include both oriented outputs of every edge in the first $d-2$
trees. For an edge $e=(i,j;D)$, the chain rule gives
\begin{align*}
 \partial_sU^s_{i\mid D\cup\{j\}}
 ={}&\nabla_{\theta_e}h_e^{\rightarrow}
       \mathbin{\cdot}(\widehat\theta_e-\theta_e^\circ)\\
 &+\partial_u h_e^{\rightarrow}\,\partial_sU^s_{i\mid D}
  +\partial_v h_e^{\rightarrow}\,\partial_sU^s_{j\mid D},
\end{align*}
where the derivatives are evaluated at the current inputs and parameter.
The reverse orientation uses $h_e^{\leftarrow}$ with the input roles exchanged.
Since the original uniforms are fixed along the path,
Equation~\eqref{eq:vine_uniform_recursion} and induction imply
\begin{equation*}
 |\partial_sU^s_{i\mid D}|\le r_{i\mid D}(\tau)
\end{equation*}
whenever all inputs in the corresponding recursion belong to
$[\tau,1-\tau]$. On the set where every vine input belongs to this
interval, differentiating all pair-density factors therefore gives
\begin{equation}
 |\partial_s\log c_s|
 \le\sum_{e=(i,j;D)\in\mathcal E}
 L_e(\tau)\left[\varepsilon_e
       +r_{i\mid D}(\tau)+r_{j\mid D}(\tau)\right].
 \label{eq:vine_local_path_score}
\end{equation}

\emph{Localisation of the square-root density.}
Choose $\varphi_\tau\in C^1([0,1];[0,1])$ such that
\begin{equation*}
 \varphi_\tau=1\ \text{on }[2\tau,1-2\tau],\qquad
 \varphi_\tau=0\ \text{on }[0,\tau]\cup[1-\tau,1],\qquad
 |\varphi_\tau'|\le2/\tau.
\end{equation*}
Define
\begin{align*}
 F_s^\tau(u)
 ={}&\sqrt{c_s(u)}\prod_{\ell=1}^d\varphi_\tau(u_\ell)\\
 &\times\prod_{k=1}^{d-2}\prod_{e=(i,j;D)\in\mathcal E_k}
 \left[\varphi_\tau(U^s_{i\mid D\cup\{j\}}(u))
       \varphi_\tau(U^s_{j\mid D\cup\{i\}}(u))\right].
\end{align*}
For $d=2$, the product over the trees equals one.
Under the probability density $c_s$, the original uniforms and every
conditional probability integral transform in this product have uniform
marginals. The total number of factors is
$d+2\sum_{k=1}^{d-2}(d-k)=d^2-2$.
Whenever every factor has its argument in $[2\tau,1-2\tau]$, the
product equals one. The union bound thus gives
\begin{equation}
 \|\sqrt{c_s}-F_s^\tau\|_2^2
 \le4(d^2-2)\tau\le4d^2\tau,
 \label{eq:vine_cutoff_mass}
\end{equation}
where the norm is with respect to Lebesgue measure on $(0,1)^d$.
This estimate uses the conditional transforms and the integrating law
from the same interpolating model.

We next differentiate $F_s^\tau$ in $s$. In every nonzero term of its
product derivative, the supports of $\varphi_\tau$ and
$\varphi_\tau'$ place all the truncated variables in $[\tau,1-\tau]$.
These variables include all recursive inputs for every density factor
and every conditional transform. Hence
Equation~\eqref{eq:vine_local_path_score} and the recursive derivative
bounds apply to each such term. Since the original uniforms have zero
parameter-path derivative, we obtain
\begin{align*}
 |\partial_sF_s^\tau|
 \le\sqrt{c_s}\Bigg\{
 &\frac12\sum_{e=(i,j;D)\in\mathcal E}
   L_e(\tau)\left[\varepsilon_e
        +r_{i\mid D}(\tau)+r_{j\mid D}(\tau)\right]\\
 &+\frac2\tau\sum_{k=1}^{d-2}
        \sum_{e=(i,j;D)\in\mathcal E_k}
   \left[r_{i\mid D\cup\{j\}}(\tau)
        +r_{j\mid D\cup\{i\}}(\tau)\right]\Bigg\}.
\end{align*}
As $\int c_s=1$, the definition in
Equation~\eqref{eq:vine_error_budget} yields
$\|\partial_sF_s^\tau\|_2\le\Delta_t(H;\tau)/2$.
Pointwise continuous differentiability along the parameter path and
this integrable norm bound justify the fundamental theorem of calculus
and Minkowski's inequality. Together with
Equation~\eqref{eq:vine_cutoff_mass}, they give
\begin{align*}
 h(c_0,c_1)
 &\le\|\sqrt{c_0}-F_0^\tau\|_2
     +\|F_0^\tau-F_1^\tau\|_2
     +\|F_1^\tau-\sqrt{c_1}\|_2\\
 &\le4d\sqrt\tau+\int_0^1\|\partial_sF_s^\tau\|_2\,ds\\
 &\le\frac12\{\Delta_t(H;\tau)+8d\sqrt\tau\}.
\end{align*}
Taking the infimum over $0<\tau<1/4$ gives $h(c_0,c_1)\le\Delta_t(H)/2$.
The common marginal change of variables in
Equation~\eqref{eq:vine_reference_model} preserves Hellinger distance,
so the same bound holds for $q_t^\circ$ and $\widehat q_t$.
Finally, Cauchy--Schwarz gives
\begin{equation*}
 \|P-Q\|_{\mathrm{TV}}
 =\frac12\int|\sqrt p-\sqrt q|(\sqrt p+\sqrt q)
 \le h(P,Q),
\end{equation*}
since $\int(\sqrt p+\sqrt q)^2\le4$.
Together with $\|P-Q\|_{\mathrm{TV}}\le1$, this proves
Equation~\eqref{eq:vine_density_stability}.

\emph{Explicit bounds in terms of dimension.}
Under Assumption~\ref{ass:vine_approximation_budget}, define
\begin{equation*}
 b_0(\tau)=0,\qquad
 b_j(\tau)=\frac A\tau\sum_{m=0}^{j-1}
                    \left(\frac{2A}{\tau}\right)^m,\qquad j\ge1.
\end{equation*}
Because $b_j=A/\tau+(2A/\tau)b_{j-1}$, the recursion implies
$r_{i\mid D}(\tau)\le\varepsilon b_{|D|}(\tau)$.
Set $x=2A/\tau>8$. Then
\begin{equation*}
 b_j(\tau)=\frac12\sum_{\ell=1}^j x^\ell
 \le\frac{x^j}{2(1-x^{-1})}\le\frac47x^j.
\end{equation*}
Tree $k$ has $d-k$ edges, each contributing at most
$\varepsilon b_k$ to the density-score sum. For $k\le d-2$, each of its
two oriented outputs has derivative bound at most $\varepsilon b_k$.
Thus
\begin{equation*}
 \Delta_t(H;\tau)
 \le\varepsilon\left[
       \sum_{k=1}^{d-1}(d-k)b_k(\tau)
       +\frac8\tau\sum_{k=1}^{d-2}(d-k)b_k(\tau)\right].
\end{equation*}
The two weighted geometric sums satisfy
\begin{align*}
 \sum_{k=1}^{d-1}(d-k)b_k(\tau)
 &\le\frac47x^{d-1}\sum_{\ell=0}^\infty(\ell+1)8^{-\ell}
 \le x^{d-1},\\
 \sum_{k=1}^{d-2}(d-k)b_k(\tau)
 &\le\frac47x^{d-2}\sum_{\ell=0}^\infty(\ell+2)8^{-\ell}
 \le\frac32x^{d-2}.
\end{align*}
The second finite sum is zero for $d=2$.
Using $A\ge1$ therefore yields
\begin{align*}
 \Delta_t(H;\tau)
 &\le\varepsilon\left(x^{d-1}+\frac{12}{\tau}x^{d-2}\right)\\
 &=\left(1+\frac6A\right)\varepsilon x^{d-1}
 \le7(2A)^{d-1}\varepsilon\tau^{-(d-1)}.
\end{align*}
For $0<\varepsilon<1$, choose
$\tau=\frac14\varepsilon^{2/(2d-1)}$ in the final error budget.
This gives
\begin{equation*}
 \Delta_t(H)
 \le\{7(8A)^{d-1}+4d\}\varepsilon^{1/(2d-1)}
 \le8^dA^{d-1}\varepsilon^{1/(2d-1)},
\end{equation*}
where $4d\le8^{d-1}$ for $d\ge2$ and $A\ge1$.
For $\varepsilon=1$, let $\tau\uparrow1/4$ in the same upper bound.
For $\varepsilon=0$, every recursive error vanishes and
$\tau\downarrow0$ gives $\Delta_t(H)=0$.
These estimates establish Equation~\eqref{eq:vine_uniform_budget}
with $C_d=8^dA^{d-1}$ throughout $0\le\varepsilon\le1$.

\end{proof}

\phantomsection\label{app:vine_model_error_proof}
\begin{proof}[Proof of Lemma~\ref{lem:vine_model_error}]

Fix a date $t$ and history $H$, and write
$q=q_t(\cdot\mid H)$, $q^\circ=q_t^\circ(\cdot\mid H)$,
$\widehat q=\widehat q_t(\cdot\mid H)$ and $\Delta=\Delta_t(H)$.
Pinsker's inequality and $h^2(P,Q)\le D_{\mathrm{KL}}(P\Vert Q)$,
applied to the budget in Assumption~\ref{ass:vine_approximation_budget},
give
\begin{equation*}
 \|q-q^\circ\|_{\mathrm{TV}}\le\frac\zeta2,
 \qquad h(q,q^\circ)\le\frac\zeta{\sqrt2}.
\end{equation*}
Theorem~\ref{thm:vine_density_stability} gives
$\|q^\circ-\widehat q\|_{\mathrm{TV}}\le\Delta/2$ and
$h(q^\circ,\widehat q)\le\Delta/2$.
The triangle inequality for each distance therefore yields
\begin{align*}
 \|q-\widehat q\|_{\mathrm{TV}}
 &\le\min\left\{1,\frac{\zeta+\Delta}{2}\right\},\\
 h(q,\widehat q)
 &\le\frac\zeta{\sqrt2}+\frac\Delta2
 \le\frac{\zeta+\Delta}{\sqrt2}.
\end{align*}
Substituting $\Delta\le C_d\varepsilon^{1/(2d-1)}$ from
Equation~\eqref{eq:vine_uniform_budget} proves
Equation~\eqref{eq:vine_total_law_error}.
\end{proof}

\phantomsection\label{app:vine_performance_proof}
\begin{proof}[Proof of Theorem~\ref{thm:vine_policy_performance}]
Fix the fitted conditional kernels and the auxiliary kernels in Assumption~\ref{ass:vine_approximation_budget}. Let $\mathbb P_{[t]}$, $\mathbb P^\circ_{[t]}$ and $\widehat{\mathbb P}_{[t]}$ denote the laws of the raw history $H_t$ obtained by iterating the true, auxiliary and fitted kernels, respectively. Their initial laws are the same point mass. Market returns are exogenous, so these history laws are independent of the investment policy. Write
\begin{equation*}
d_t(H)=\|q_t(\cdot\mid H)-\widehat q_t(\cdot\mid H)\|_{\mathrm{TV}}.
\end{equation*}
The true-to-auxiliary KL budget in Assumption~\ref{ass:vine_approximation_budget} implies $q_r(\cdot\mid H)\ll q_r^\circ(\cdot\mid H)$. The entropy chain rule therefore gives
\begin{align}
D_{\mathrm{KL}}(\mathbb P_{[t]}\|\mathbb P^\circ_{[t]})
&=\sum_{r<t}\mathbb E\!\left[
D_{\mathrm{KL}}\bigl(q_r(\cdot\mid H_r)\|q_r^\circ(\cdot\mid H_r)\bigr)\right]\notag\\
&\leq\frac{t\zeta^2}{2}.
\label{eq:vine_path_entropy}
\end{align}
By Pinsker's inequality and $h(P,Q)^2\leq D_{\mathrm{KL}}(P\Vert Q)$, Equation~\eqref{eq:vine_path_entropy} yields
\begin{equation*}
\|\mathbb P_{[t]}-\mathbb P^\circ_{[t]}\|_{\mathrm{TV}}
\leq\frac{\sqrt t\,\zeta}{2},
\qquad
h(\mathbb P_{[t]},\mathbb P^\circ_{[t]})\leq\sqrt{t/2}\,\zeta.
\end{equation*}
To compare the auxiliary and fitted path laws, write their densities with respect to a common dominating measure $\mu_{[t]}$ as $p^\circ_{[t]}$ and $\widehat p_{[t]}$. Factoring the one-step path densities in the Hellinger affinity gives
\begin{align*}
&h^2(\mathbb P^\circ_{[t+1]},\widehat{\mathbb P}_{[t+1]})\\
&\quad=h^2(\mathbb P^\circ_{[t]},\widehat{\mathbb P}_{[t]})\\
&\qquad+\int\sqrt{p^\circ_{[t]}\widehat p_{[t]}}\,
h^2\bigl(q_t^\circ(\cdot\mid H),\widehat q_t(\cdot\mid H)\bigr)\,d\mu_{[t]}(H)\\
&\quad\leq h^2(\mathbb P^\circ_{[t]},\widehat{\mathbb P}_{[t]})
+\frac{C_d^2\varepsilon^{2/(2d-1)}}4.
\end{align*}
Here we used Theorem~\ref{thm:vine_density_stability} and
$\int\sqrt{p^\circ_{[t]}\widehat p_{[t]}}\,d\mu_{[t]}\leq1$.
Iteration from the common initial law gives
\begin{equation*}
h(\mathbb P^\circ_{[t]},\widehat{\mathbb P}_{[t]})
\leq\frac{C_d\sqrt t}{2}\varepsilon^{1/(2d-1)}.
\end{equation*}
The triangle inequality and $\|P-Q\|_{\mathrm{TV}}\leq h(P,Q)$ combine the two path comparisons. Together with the conditional bound in Lemma~\ref{lem:vine_model_error}, they give
\begin{align}
\|\mathbb P_{[t]}-\widehat{\mathbb P}_{[t]}\|_{\mathrm{TV}}
&\leq\frac{\sqrt t}{2}(\zeta+C_d\varepsilon^{1/(2d-1)}),\notag\\
\mathbb E[d_t(H_t)]
&\leq\frac{\zeta+C_d\varepsilon^{1/(2d-1)}}{2}.
\label{eq:vine_path_tv}
\end{align}
The Hellinger triangle inequality also gives
\begin{align}
h(\mathbb P_{[t]},\widehat{\mathbb P}_{[t]})
&\leq\sqrt t\left(\frac\zeta{\sqrt2}
+\frac{C_d\varepsilon^{1/(2d-1)}}2\right)\notag\\
&\leq\sqrt{t/2}\,(\zeta+C_d\varepsilon^{1/(2d-1)}).
\label{eq:vine_path_hellinger}
\end{align}

We next compare the conditional risk charges. For a bounded measurable function $f$ and probability measures $\mu,\nu$, our total variation convention gives
\begin{equation}
|\mathbb E_\mu f-\mathbb E_\nu f|
\leq\operatorname{osc}(f)\|\mu-\nu\|_{\mathrm{TV}},
\qquad \operatorname{osc}(f)=\sup f-\inf f.
\label{eq:vine_tv_expectation}
\end{equation}
Fix $t<T$, a state $s=(H,X,b)\in\mathcal S_t$ and an action $a\in A_t(s)$. The proof of Proposition~\ref{prop:dynamic_programming} restricts both minimisations in Equation~\eqref{eq:conditional_cvar} to the common interval $I_v$ in Equation~\eqref{eq:cvar_threshold_interval} and bounds the hinge integrand uniformly by $(B-\underline k)/\alpha$ on that interval.
The difference of the two minima is bounded by the uniform difference of their threshold objectives on $I_v$. The common term $v$ cancels, and Equation~\eqref{eq:vine_tv_expectation} yields
\begin{align}
|\mathcal C_t(s,a)-\widehat{\mathcal C}_t(s,a)|
&\leq\frac1p\sup_{v\in I_v}
\left|\int_{\mathcal G}(\ell_t(s,a,g)-v)_+
\bigl[q_t(dg\mid H)-\widehat q_t(dg\mid H)\bigr]\right|\notag\\
&\leq\frac{B-\underline k}{\alpha p}\,d_t(H).
\label{eq:vine_conditional_cvar_error}
\end{align}

Fix $\pi\in\Pi$. Starting from the common initial state, recursively apply the policy and the physical update $\Phi_t$ in Equation~\eqref{eq:ideal_market_dynamics} along a given return history. The policy is Borel measurable and $\Phi_t$ is continuous, so this construction defines measurable functions $s_t^\pi(H_t)$ and $a_t^\pi(H_t)$ that are identical under both models. Admissibility keeps these states and actions feasible at every history. Put
\begin{align*}
C_t^\pi(H)
&=\mathcal C_t\bigl(s_t^\pi(H),a_t^\pi(H)\bigr),\\
\widehat C_t^\pi(H)
&=\widehat{\mathcal C}_t\bigl(s_t^\pi(H),a_t^\pi(H)\bigr).
\end{align*}
By Assumption~\ref{ass:market}, both functions take values in $[1-B,1-\underline k]$. We add and subtract $\mathbb E\widehat C_t^\pi(H_t)$, then apply Equations~\eqref{eq:vine_conditional_cvar_error}, \eqref{eq:vine_tv_expectation} and \eqref{eq:vine_path_tv} to obtain
\begin{align}
\left|\mathbb E C_t^\pi(H_t)
      -\widehat{\mathbb E}\widehat C_t^\pi(H_t)\right|
&\leq\mathbb E
       |C_t^\pi(H_t)-\widehat C_t^\pi(H_t)|\notag\\
&\quad+
       |\mathbb E\widehat C_t^\pi(H_t)
        -\widehat{\mathbb E}\widehat C_t^\pi(H_t)|\notag\\
&\leq\frac{B-\underline k}{\alpha p}\mathbb E[d_t(H_t)]
       +(B-\underline k)\|\mathbb P_{[t]}-\widehat{\mathbb P}_{[t]}\|_{\mathrm{TV}}\notag\\
&\leq\frac{B-\underline k}{2}(\zeta+C_d\varepsilon^{1/(2d-1)})
       \left(\frac1{\alpha p}+\sqrt t\right).
\label{eq:vine_cvar_path_error}
\end{align}
The first term compares risk charges at the same realised history and portfolio state. The second term controls the change in the distribution of that history and the portfolio state it determines.

We next compare the expected terminal utilities using their common second-moment bound in Equation~\eqref{eq:vine_terminal_moment}. For probability measures $P,Q$ dominated by $\mu$, write $\rho=dP/d\mu$ and $\sigma=dQ/d\mu$. For $f\in L^2(P)\cap L^2(Q)$, the square-root factorisation of the density difference and Cauchy--Schwarz give
\begin{align*}
|\mathbb E_P f-\mathbb E_Q f|
&=\left|\int f(\sqrt\rho-\sqrt\sigma)(\sqrt\rho+\sqrt\sigma)\,d\mu\right|\\
&\leq h(P,Q)\left[\int f^2(\sqrt\rho+\sqrt\sigma)^2\,d\mu\right]^{1/2}\\
&\leq h(P,Q)\sqrt{2(\mathbb E_P f^2+\mathbb E_Q f^2)}.
\end{align*}
We apply this estimate to the common terminal-utility function $f=U_{\gamma_U}(X_T^\pi)$ under $P=\mathbb P_{[T]}$ and $Q=\widehat{\mathbb P}_{[T]}$. Equations~\eqref{eq:vine_terminal_moment} and \eqref{eq:vine_path_hellinger} give
\begin{align}
\left|\mathbb E U_{\gamma_U}(X_T^\pi)
      -\widehat{\mathbb E}U_{\gamma_U}(X_T^\pi)\right|
&\leq2\sqrt{M_T}\,h(\mathbb P_{[T]},\widehat{\mathbb P}_{[T]})\notag\\
&\leq\sqrt{2TM_T}\,(\zeta+C_d\varepsilon^{1/(2d-1)}).
\label{eq:vine_terminal_utility_error}
\end{align}
The utility increments in Equation~\eqref{eq:investment_objective} telescope along each return path, so
\begin{align*}
J(\pi;s_0)
&=\mathbb E U_{\gamma_U}(X_T^\pi)-U_{\gamma_U}(X_0)
-\lambda\sum_{t<T}\mathbb E C_t^\pi(H_t),\\
\widehat J(\pi;s_0)
&=\widehat{\mathbb E} U_{\gamma_U}(X_T^\pi)-U_{\gamma_U}(X_0)
-\lambda\sum_{t<T}\widehat{\mathbb E}\widehat C_t^\pi(H_t).
\end{align*}
Combining Equations~\eqref{eq:vine_cvar_path_error} and \eqref{eq:vine_terminal_utility_error}, and using $\sum_{t=0}^{T-1}\sqrt t\leq\int_0^T\sqrt s\,ds=\tfrac23T^{3/2}$, yields
\begin{equation*}
|J(\pi;s_0)-\widehat J(\pi;s_0)|
\leq\mathcal K_T(\zeta+C_d\varepsilon^{1/(2d-1)}),
\end{equation*}
with $\mathcal K_T$ defined in Equation~\eqref{eq:vine_performance_bound}.
The bound is uniform over $\pi\in\Pi$, proving the first assertion in Equation~\eqref{eq:vine_value_stability}. The moment bound and bounded risk charges also make both optimal values finite. Since $\Pi$ is nonempty,
\begin{equation*}
|V^\Pi-\widehat V^\Pi|
\leq\sup_{\pi\in\Pi}|J(\pi;s_0)-\widehat J(\pi;s_0)|
\leq\mathcal K_T(\zeta+C_d\varepsilon^{1/(2d-1)}),
\end{equation*}
which proves the second assertion.

For logarithmic utility, we verify Equation~\eqref{eq:vine_log_moment} directly from the market conditions. Along any admissible policy, write $K_t=K_t(s_t^\pi,a_t^\pi,G_{t+1})$ and $\log^+x=\max\{0,\log x\}$. The elementary bound $(\log^+x)^2\leq4x/e^2$, the lower bound $K_t\geq\underline k$ and the conditional first-moment bound imply
\begin{equation*}
\mathbb E\bigl[|\log K_t|^2\mid H_t\bigr]
\leq(\max\{0,-\log\underline k\})^2+\frac{4B}{e^2}.
\end{equation*}
The same estimate holds with $\widehat{\mathbb E}$. Since $X_0=1$, the wealth recursion and Cauchy--Schwarz give
\begin{equation*}
|\log X_T^\pi|^2
=\left|\sum_{t=0}^{T-1}\log K_t\right|^2
\leq T\sum_{t=0}^{T-1}|\log K_t|^2.
\end{equation*}
Taking expectations under each model proves Equation~\eqref{eq:vine_log_moment} uniformly over admissible policies.
\end{proof}

\phantomsection\label{app:vine_transfer_proof}
\begin{proof}[Proof of Theorem~\ref{thm:vine_policy_transfer}]
Since $\widehat\pi\in\Pi$, its true value is at most $V^\Pi$. By Theorem~\ref{thm:vine_policy_performance} and the fitted-model optimisation guarantee,
\begin{align*}
V^\Pi-J(\widehat\pi;s_0)
&=(V^\Pi-\widehat V^\Pi)
  +(\widehat V^\Pi-\widehat J(\widehat\pi;s_0))\notag\\
&\quad+(\widehat J(\widehat\pi;s_0)-J(\widehat\pi;s_0))\notag\\
&\leq2\mathcal K_T(\zeta+C_d\varepsilon^{1/(2d-1)})+\eta.
\end{align*}
This proves Equation~\eqref{eq:vine_transfer_bound}.
\end{proof}

\phantomsection\label{app:vine_learning_proof}
\begin{proof}[Proof of Corollary~\ref{cor:vine_learning}]
On the event in Equation~\eqref{eq:vine_learning_conditions}, the policy-uniform simulation error and the empirical optimisation guarantee give
\begin{align*}
\widehat V^\Pi-\widehat J(\widehat\pi;s_0)
&\leq\sup_{\pi\in\Pi}J_{\mathrm{sim}}(\pi)-J_{\mathrm{sim}}(\widehat\pi)+2E_{\mathrm{sim}}\\
&\leq\eta_{\mathrm{opt}}+2E_{\mathrm{sim}}.
\end{align*}
Applying Theorem~\ref{thm:vine_policy_transfer} with $\eta=2E_{\mathrm{sim}}+\eta_{\mathrm{opt}}$ proves Equation~\eqref{eq:vine_learning_bound}. If model fitting or policy training is random, we apply this argument separately to each realisation satisfying the stated event.
\end{proof}


{\emergencystretch=2em
\footnotesize
\printbibliography
}

\end{document}